\documentclass[12pt,fleqn]{article} \textheight=9in
\newcommand{\ra}{\rightarrow}
\newcommand{\bra}{\langle} \newcommand{\ket}{\rangle}
\newcommand{\be}{\begin{equation}}
\newcommand{\ee}{\end{equation}}
\newcommand{\bea}{\begin{eqnarray}}
\newcommand{\eea}{\end{eqnarray}}
\newcommand{\eps}{\epsilon}
\newcommand{\E}{\mbox{e}}
\newcommand{\e}{\mbox{\scriptsize e}}
\newcommand{\ffi}{\varphi}

\newcommand{\ep}{\qquad {\vrule height 10pt width 8pt depth 0pt}}

\newcommand{\grintl}{[\kern-.18em [}
\newcommand{\grintr}{]\kern-.18em ]}
\newcommand{\ds}{\displaystyle}
\newtheorem{lem}{Lemma}[section]
\newtheorem{prop}{Proposition}[section]
\newtheorem{thm}{Theorem}[section]

\def\smallR{\hbox{\scriptsize I\kern-.23em{R}}}
\def\R{\hbox{$\mit I$\kern-.33em$\mit R$}}
\def\C{\hbox{$\mit I$\kern-.6em$\mit C$}}
\def\un{\hbox{$\mit I$\kern-.77em$\mit I$}}
\def\0{\hbox{$\mit I$\kern-.70em$\mit O$}}
\def\r{I\kern-.277em R}

\def\N{\mbox{\bf N}}
\def\dist{\mbox{\rm dist}}

\begin{document}

\title{A Time--Dependent Born--Oppenheimer Approximation with Exponentially
Small Error Estimates}
\author{George A. Hagedorn\thanks{Partially
Supported by National Science Foundation
Grant DMS--9703751.}\\
Department of Mathematics and\\
Center for Statistical Mechanics and Mathematical Physics\\
Virginia Polytechnic Institute and State University\\
Blacksburg, Virginia 24061-0123, U.S.A.\\[15pt]
\and
Alain Joye\\
Institut Fourier\\ Unit\'e Mixte de Recherche CNRS-UJF 5582\\
Universit\'e de Grenoble I\\
BP 74\\
F--38402 Saint Martin d'H\`eres Cedex, France}

\date{1 May 2000}
\maketitle

\begin{abstract}
We present the construction of an exponentially accurate
time--dependent Born--Oppenheimer approximation for molecular quantum
mechanics.

We study molecular systems whose electron masses are held fixed and whose
nuclear masses are proportional to $\epsilon^{-4}$, where $\epsilon$ is a
small expansion parameter. By optimal truncation of an asymptotic
expansion, we construct approximate solutions to the time--dependent
Schr\"odinger equation that agree with exact normalized solutions up to errors
whose norms are bounded by $\ds C\,\exp\left(\,-\gamma/\epsilon^2\,\right)$, for
some $C$ and $\gamma>0$. \end{abstract}

\newpage

\section{Introduction}\label{intro}
\setcounter{equation}{0}

\vskip .5cm
In this paper we construct exponentially accurate approximate solutions to
the time--de\-pen\-dent Schr\"odinger equation for a molecular system. The
small parameter that governs the approximation is the usual
Born--Oppenheimer expansion parameter $\epsilon$, where $\epsilon^4$ is
the ratio of the electron mass divided by the mean nuclear mass.
The approximate solutions we construct agree with exact solutions up to
errors whose norms are bounded by
$\ds C\,\exp\left(\,-\gamma/\epsilon^2\,\right)$,
for some $C$ and $\gamma>0$, under analyticity assumptions on the 
electron Hamiltonian.

The Hamiltonian for a molecular system with $K$ nuclei and $N-K$
electrons moving in $l$ dimensions has the form
$$H(\epsilon)\ =\
\sum_{j=1}^K\,-\,\frac{\epsilon^4}{2M_j}\,\Delta_{X_j}\,-\,
\sum_{j=K+1}^N\,\frac{1}{2m_j}\,\Delta_{X_j}\,+\,
\sum_{i<j}\,V_{ij}(X_i-X_j).
$$
Here $X_j\in\R^l$ denotes the position of the $j^{\mbox{\scriptsize th}}$
particle, the mass of the $j^{\mbox{\scriptsize th}}$ nucleus is
$\epsilon^{-4}M_j$ for $1\le j\le K$, the mass of the
$j^{\mbox{\scriptsize th}}$ electron is $m_j$ for $K+1\le j\le N$, and
the potential between particles $i$ and $j$ is $V_{ij}$. For convenience,
we assume each $M_j=1$. We set $d=Kl$ and let
$X=(X_1,\,X_2,\,\dots,\,X_K)\in\R^d$ denote the nuclear configuration
vector. We can then decompose $H(\epsilon)$ as
$$
H(\epsilon)\ =\ -\,\frac{\epsilon^4}{2}\,\Delta_X\,+\,h(X).
$$
The first term on the right hand side
represents the nuclear kinetic energy, and the second
is the ``electron Hamiltonian'' that depends parametrically on $X$. For
each fixed $X$, $h(X)$ is a self-adjoint operator on the Hilbert space
${\cal H}_{\mbox{\scriptsize el}}=L^2(\R^{(N-K)l})$.

The time--dependent Schr\"odinger equation we approximately solve in
$L^2(\R^d,\, {\cal H}_{\mbox{\scriptsize el}})$ as $\epsilon \ra 0$ is
$$
i\,\epsilon^2\,\frac{\partial\psi}{\partial t}\ =\
-\,\frac{\epsilon^4}{2}\,\Delta_X\,\psi\,+\,h(X)\,\psi
$$
Asymptotic expansions in powers of $\epsilon$ of certain solutions to this
equation are derived in \cite{oldstuff1,oldstuff2,oldstuff3}. We obtain our
construction by truncating these expansions after an $\epsilon$--dependent
number of terms, in an effort to minimize the
norm of the error. Similar strategies have been used to obtain exponentially
accurate results for adiabatic
approximations \cite{nenciu, alainetal1, alainetal2, j} and semiclassical approximations
\cite{hagjoy3, hagjoy5}, both of which play roles in the Born--Oppenheimer
approximation we are studying here.

Roughly speaking, the time--dependent Born--Oppenheimer approximation says
the following for small $\epsilon$: The electrons move very rapidly and
adjust their state adiabatically as the more slowly moving nuclei change
their positions. If the electrons start in a discrete energy level of $h(X)$,
they will remain in that level. In the process, the electron states create an
effective potential in which the motion of the heavy nuclei is well
described by a semiclassical approximation. The asymptotic expansions show
that this intuition is valid up to errors of order $\epsilon^k$ for any
$k$.

In Born--Oppenheimer approximations, adiabatic and semiclassical limits
are being taken simultaneously, and they are coupled.
Analysis of errors for the adiabatic and semiclassical approximations
shows that they are each accurate up to errors whose bounds have the form
$C\,\exp\left(\,-\,\gamma/\eps^2\,\right)$ \cite{jp, hagjoy3}. 
Non-adiabatic transitions are
known in some systems to be of this order, and tunnelling in semiclassical
approximations makes contributions of this order. Thus, one cannot expect to do
better than approximations of this type because of two well-known physical
phenomena that Born--Oppenheimer approximations do not take into account.

In some systems, tunnelling might dominate the error. In some,
non-adiabatic electronic transitions may dominate. In others, the two effects
can be of comparable magnitude.

One of the motivations
for our work is to generate a ``good'' basis upon which to build a
``surface hopping model'' that would accurately describe non-adiabatic
electronic transitions. Prior authors (see, {\it e.g.},
\cite{pechukas,tully,webster,coker})
have proposed such models based on the zeroth order
time--dependent Born--Oppenheimer approximation. Using the
zeroth order states as a
basis of the surface hopping model, the non-adiabatic transitions appear
at order $\epsilon^2$. This is huge compared to the exponentially small
physical phenomenon one would like to study, and we believe interference between
transitions that occur at different times is responsible for the exponential
smallness of the physically interesting quantity.
Our view is that by choosing
a much better set of states on which to base the model, one will obtain a
much more useful approximation. Sir Michael Berry \cite{lots!a, lots!b, lots!2} has
advocated such ideas for the somewhat simpler adiabatic approximation
(which does not have the complications of the nuclear motion). These ideas have
been used in \cite{j,alainetal1} to prove the accuracy of certain results for
non--adiabatic transitions that are exponentially small.

\vskip .3cm
\noindent {\bf Remarks:}

\vskip .1cm
\noindent {\bf 1.}\quad There are some other exponentially accurate results
in the general topic of Born--Oppenheimer approximations. The prior results
come from study of the time--independent Schr\"odinger equation and depend on
global properties of the system.
Our results are time--dependent and make use of local information.

Klein \cite{h30} and Martinez \cite{h34,h36,h37} show that resonances associated
with predissociation processes have exponentially long lifetimes. Benchaou and
Martinez \cite{benchaou,benmar} also show that certain S--matrix elements
associated with non--adiabatic transitions are exponentially small.

\vskip .3cm
\noindent {\bf 2.}\quad The papers cited in the previous remark obtain estimates
that depend on the global structure of the electron energy levels.
The results we obtain depend on a particular classical path. When the path
stays away from the nuclear configurations where the gap between relevant
electonic levels is minimized, one would expect the non--adiabatic errors
from our
approximation to be smaller, {\it i.e.}, both results would obtain errors
of order $\exp(-\Gamma/\eps^2)$, but we would obtain a larger value of
$\Gamma$.

We expect this because in our case, the Landau--Zener formula predicts that
our $\Gamma$ should come from the minimum gap between eigenvalues on the
classical path, rather than the global minimum gap.

\vskip .3cm
\noindent {\bf 3.}\quad From a mathematical point of view, the optimal
truncation procedure in this context was first stated for the adiabatic 
approximation for two component systems
of ODE's by Berry \cite{lots!a, lots!b}. It was first proved to yield exponentially
accurate results for Hilbert space valued ODE's by Nenciu \cite{nenciu}.
In \cite{hagjoy3,hagjoy5} we used this idea for the semiclassical approximation,
which is a complex valued PDE setting. The present paper can be viewed as
extending these ideas to a Hilbert space valued PDE setting.

\vskip .5cm
\subsection{Hypotheses}

We assume that the electron Hamiltonian $h(X)$ satisfies the following
analyticity hypotheses:

\begin{itemize}
\item [{\bf H$_0$}]
\begin{itemize}
\item[{\bf (i)}] For any $X\in\R^d$, $h(X)$ is a self-adjoint operator on some dense domain
${\cal D}\subset{\cal H}_{\mbox{\scriptsize el}}$,
where ${\cal H}_{\mbox{\scriptsize el}}$ is the electronic Hilbert
space. We assume the domain ${\cal D}$ is independent of $X$ and $h(X)$ is
bounded from below uniformly in $\R^d$.
\item[{\bf (ii)}] There exists a $\delta>0$, such that
for every $\psi\in{\cal D}$, the vector $h(X)\psi$ is
analytic in\\ $S_{\delta}=\{z\in\C^d\; :\; |\mbox{Im}(z_j)|<\delta,\;\;
j=1,\dots, d\}$.
\end{itemize}
\item[{\bf H$_1$}]
There exists an open set $\Xi\subset\R^d$, such that for all $X\in\Xi$,
there exists an isolated, multiplicity one eigenvalue $E(X)$ of
$h(X)$ associated with a normalized eigenvector
$\Phi(X)\in {\cal H}_{\mbox{\scriptsize el}}$. We assume without
loss that the origin belongs to $\Xi$.
\end{itemize}

\vspace{.3cm}
\noindent {\bf Remarks:}\quad
1.\quad Hypothesis {\bf H$_0$}  implies that the family of
operators $\{h(X)\}_{X\in S_{\delta}}$ is a holomorphic family
of type A.

2.\quad It follows from  {\bf H$_0$} and {\bf H$_1$} that there exists
$\delta'\in(0,\,\delta)$ and $\Xi'\subset \Xi$ such that the complex and vector valued
functions $E(\cdot)$ and $\Phi(\cdot)$ admit analytic continuations
on the set\\
$\Sigma_{\delta'}=\{z\in\C^d\,:\,\mbox{Re}(z)\in\Xi'\quad\mbox{and}\quad
|\mbox{Im}(z_j)|<\delta',\;\; j=1,\dots,d\}$.

\vskip .5cm
\subsection{Summary of the Main Results}

Our main results are stated precisely as Theorem \ref{main} in Section \ref{mr}.
Two generalizations of this result are presented in Section \ref{gene}. 

\vskip .2cm
Roughly speaking, Theorem \ref{main} states the following:

\vskip .5cm
Under hypotheses
{\bf H}$_0$ and {\bf H}$_1$, we construct $\Psi_*(X,t,\eps)$ (that depends on
a parameter $g$) for $t\in [0,\,T]$. For small values of $g$, there exist $C(g)$
and $\Gamma(g)>0$, such that in the limit
$\eps\ra 0$,
$$
\left\|\,\E^{-itH(\eps)/\eps^2}\Psi_*(X,0,\eps)\,-\,\Psi_*(X,t,\eps)\,
\right\|_{L^2({\smallR}^d,{\cal H}_{\mbox{\scriptsize el}})}\
\leq\ C(g)\ \E^{-\Gamma(g)/\eps^2}
$$
In the state $\Psi_*(X, t, \eps)$, the electrons have a high probability of
being in the electron state $\Phi(X)$. For any $b>0$ and sufficiently small
values of $g$, the nuclei are localized near a classical path $a(t)$ in the
sense that there exist $c(g)$ and $\gamma(g)>0$, such that in the limit
$\eps\ra 0$,
$$
\left(\,\int_{|X-a(t)|>b}\,
\|\Psi_*(X, t, \eps)\|^2_{{\cal H}_{\mbox{\scriptsize el}}}\,
dx\,\right)^{1/2}\ \leq\  c(g)\ \E^{-\gamma(g)/\eps^2}.
$$
The mechanics of the nuclear configuration $a(t)$ is determined by classical
dynamics in the effective potential $E(X)$.

\vskip .3cm
Two theorems in Section \ref{gene} generalize this result.
The first allows the time interval to grow as $\epsilon$ tends to
zero. The second allows more general initial conditions.

\vskip .5cm
\section{Coherent States and Classical Dynamics}\label{wp}
\setcounter{equation}{0}

In the construction of our approximation to the solution of the molecular
Schr\"odinger equation, we need wave packets that describe the
semiclassical
dynamics of the heavy nuclei. In the present context, the semiclassical
parameter is $\hbar=\eps^2$. We make use of a convenient
set of coherent states (also called generalized squeezed states), that we
express here in terms of the semiclassical parameter $\hbar$.

\vskip .25cm
We recall the definition of the coherent states
$\ffi_j(A,\,B,\,\hbar,\,a,\,\eta,\,X)$ that are described in detail in
\cite{raise}. A more explicit, but more complicated definition is given in
\cite{semi4}.

We adopt the standard multi-index notation. A multi-index
$j=(j_1,\,j_2,\,\dots ,\,j_d)$ is a $d$-tuple of non-negative integers.
We define
%
%
$|j|=\sum_{k=1}^d\,j_k$,\ \,
$X^j=X_1^{j_1}X_2^{j_2}\cdots X_d^{j_d}$,\\
$j!=(j_1!)(j_2!)\cdots(j_d!)$,\ \,
and\ \,
$D^j=\frac{\partial^{|j|}}{(\partial X_1)^{j_1}(\partial X_2)^{j_2}\cdots
(\partial X_d)^{j_d}}$.

Throughout the paper we assume $a\in\R^d$, $\eta\in\R^d$ and $\hbar>0$. We
also assume that $A$ and $B$ are $d\times d$ complex invertible matrices that
satisfy
\bea\label{cond1}
A^t\,B\,-\,B^t\,A&=&0,\nonumber \\
A^*\,B\,+\,B^*\,A&=&2\,I.
\eea

These conditions guarantee that both the real and imaginary parts of $BA^{-1}$
are symmetric. Furthermore, $\mbox{Re}\,BA^{-1}$ is strictly positive definite,
and $\left(\mbox{Re}\,BA^{-1}\right)^{-1}=\,A\,A^*$.

Our definition of $\ffi_j(A,\,B,\,\hbar,\,a,\,\eta,\,X)$ is based on the
following raising operators that are defined for $m=1,\,2,\,\dots ,\,d$.
$$
\mathcal{A}_m(A,B,\hbar,a,\eta)^*
\ =\
\frac{1}{\sqrt{2\hbar}}\,\left[\,\sum_{n=1}^d\,\overline{B}_{n\,m}\,(X_n-a_n)
\ -\,i\ \sum_{n=1}^d\,\overline{A}_{n\,m}\,
(-i\hbar\frac{\partial\phantom{X^n}}{\partial X_n}-\eta_n)\,\right] .
$$
The corresponding lowering operators $\mathcal{A}_m(A,B,\hbar,a,\eta)$ are
their formal adjoints.

These operators satisfy commutation relations that lead to the properties of
the\\
$\ffi_j(A,\,B,\,\hbar,\,a,\,\eta,\,X)$ that we list below.
The raising operators $\mathcal{ A}_m(A,B,\hbar,a,\eta)^*$ for\\
$m=1,\,2,\,\dots,\,d$ commute with one another, and the lowering operators
$\mathcal{A}_m(A,B,\hbar,a,\eta)$ commute with one another. However,
$$\mathcal{A}_m(A,B,\hbar,a,\eta)\,\mathcal{A}_n(A,B,\hbar,a,\eta)^*\,-\,
\mathcal{A}_n(A,B,\hbar,a,\eta)^*\,\mathcal{A}_m(A,B,\hbar,a,\eta)
\ =\ \delta_{m,\,n}.$$

\vskip .25cm
\noindent
{\bf Definition}\quad
For the multi-index $j=0$, we define the normalized complex Gaussian wave packet
(modulo the sign of a square root) by
\bea\nonumber
&&\ffi_0(A,\,B,\,\hbar,\,a,\,\eta,\,X)\,=\,\pi^{-d/4}\,
\hbar^{-d/4}\,(\det(A))^{-1/2}\\[6pt]\nonumber
&&\qquad\qquad\quad \times\quad
\exp\left\{\,-\,\langle\,(X-a),\,B\,A^{-1}\,(X-a)\,\rangle/(2\hbar)\,
+\,i\,\langle\,\eta,\,(X-a)\,\rangle/\hbar\,\right\} .
\eea
Then, for any non-zero multi-index $j$, we define
\bea\nonumber
\ffi_j(A,\,B,\,\hbar,\,a,\,\eta,\,\cdot\,)&=&
\frac{1}{\sqrt{j!}}\
\left(\,\mathcal{A}_1(A,B,\hbar,a,\eta)^*\right)^{j_1}\
\left(\,\mathcal{A}_2(A,B,\hbar,a,\eta)^*\right)^{j_2}\ \cdots \\[5pt]
\nonumber &&\qquad\qquad\qquad\ \times\
\left(\,\mathcal{A}_d(A,B,\hbar,a,\eta)^*\right)^{j_d}\
\ffi_0(A\,,B,\,\hbar,\,a,\,\eta,\,\cdot\,).
\eea

\vskip .25cm
\noindent
{\bf Properties}\quad
1.\quad For $A=B=I$, $\hbar=1$, and $a=\eta=0$, the
$\ffi_j(A,\,B,\,\hbar,\,a,\,\eta,\,\cdot\,)$ are just the standard Harmonic
oscillator eigenstates with energies $|j|+d/2$.\\[7pt]
2.\quad For each admissible $A$, $B$, $\hbar$, $a$, and $\eta$,
the set $\{\,\ffi_j(A,\,B,\,\hbar,\,a,\,\eta,\,\cdot\,)\,\}$
is an orthonormal basis for $L^2(\R^d)$.\\[7pt]
3.\quad In \cite{semi4}, the state $\ffi_j(A,\,B,\,\hbar,\,a,\,\eta,\,X)$ is
defined as a normalization factor times
$$
\mathcal{H}_j(A;\,\hbar^{-1/2}\,|A|^{-1}\,(X-a))\
\ffi_0(A,\,B,\,\hbar,\,a,\,\eta,\,X).
$$
Here $\mathcal{H}_j(A;\,y)$ is a recursively defined
$|j|^{\mbox{\scriptsize th}}$
order polynomial in $y$ that depends on $A$ only through
$U_A$, where $A=|A|\,U_A$ is the polar decomposition of $A$.\\[7pt]
4.\quad By scaling out the $|A|$
and $\hbar$ dependence and using Remark 3 above, one can show that
$\mathcal{H}_j(A;\,y)\,\E^{-y^2/2}$ is an (unnormalized) eigenstate of the usual
Harmonic oscillator with energy $|j|+d/2$.\\[7pt]
5.\quad When the dimension $d$ is $1$, the position and momentum uncertainties
of the\newline
$\ffi_j(A,\,B,\,\hbar,\,a,\,\eta,\,\cdot\,)$ are
$\sqrt{(j+1/2)\hbar}\ |A|$ and $\sqrt{(j+1/2)\hbar}\ |B|$, respectively. In
higher dimensions, they are bounded by
$\sqrt{(|j|+d/2)\hbar}\ \|A\|$ and $\sqrt{(|j|+d/2)\hbar}\ \|B\|$,
respectively.\\[7pt]
6.\quad When we approximately solve the Schr\"odinger equation, the choice of
the sign of the square root in the definition of
$\ffi_0(A,\,B,\,\hbar,\,a,\,\eta,\,\cdot\,)$ is determined by continuity in $t$
after an arbitrary initial choice.

\vskip .5cm
The following simple but very useful lemma is proven in \cite{hagjoy5}.

\begin{lem}\label{ylem}
Let $P_{|j|\le n}$ denote the projection
onto the span of the $\ffi_j(A,\,B,\,\hbar,\,a,\,\eta,\,\cdot\,)$
with $|j|\le n$.
\be\label{simple}
(X-a)^m\ P_{|j|\le n}\ =\ P_{|j|\le n+|m|}\ (X-a)^m\ P_{|j|\le n},
\ee
and
\be\label{norm}
\left\|\,(X-a)^m\,P_{|j|\le n}\,\right\|\ \le \
\left(\,\sqrt{2\hbar}\ d\,\|\,A\,\|\,\right)^{|m|}\
\left(\,\frac{(n+|m|)!}{n!}\,\right)^{1/2}.
\ee
\end{lem}

In the Born-Oppenheimer approximation, the semiclassical dynamics
of the nuclei is generated by an effective potential given
by a chosen isolated electronic eigenvalue $E(X)$ of the electronic
hamiltonian $h(X)$, $X\in\R^d$.  For a given effective potential
$E(X)$ we describe the semiclassical dynamics of the nuclei by means
of the time dependent basis constructed as follows:

By assumption {\bf H$_1$}, the potential $E:\Xi\subset\R^d\ra\R$ is
smooth and bounded below. Associated to $E(X)$, we have the
following classical equations of motion:
\bea
\dot{a}(t)&=&\eta(t) ,\nonumber \\
\dot{\eta}(t)&=&-\,\nabla E(a(t)),\nonumber \\
\dot{A}(t)&=&i\,B(t),\label{newton} \\
\dot{B}(t)&=&i\,E^{(2)}(a(t))\,A(t),\nonumber \\
\dot{S}(t)&=&\frac{\eta(t)^2}2\,-\,E(a(t)),\nonumber
\eea
where $E^{(2)}$ denotes the Hessian matrix for $E$. We always assume the
initial conditions $A(0)$, $B(0)$, $a(0)$, $\eta(0)$, and $S(0)=0$
satisfy (\ref{cond1}).

The matrices $A(t)$ and $B(t)$ are related to the linearization of the
classical flow through the following identities:
\bea
A(t)&=&\frac{\partial a(t)}{\partial a(0)}\,A(0)\,+\,
i\,\frac{\partial a(t)}{\partial \eta(0)}\,B(0),\nonumber\\[4pt]\nonumber
B(t)&=&\frac{\partial\eta(t)}{\partial \eta(0)}\,B(0)
\,-\,i\frac{\partial\eta(t)}{\partial a(0)}\,A(0).
\eea

Because $E$ is smooth and bounded below, there exist global solutions to
the first two equations of the system (\ref{newton}) for any initial
condition if $\Xi=\R^d$. From this, it follows immediately that the
remaining three
equations of the system (\ref{newton}) have global solutions.
If $\Xi\neq\R^d$, for any initial conditions, there exists a
$0<T\leq\infty$ so that solutions to the
system (\ref{newton}) exist for any time $t\in [0,\,T]$. $T$ is finite
if and only if the solution $a(t)$ corresponding to the chosen initial
condition leaves the set $\Xi$ in finite time.

Furthermore, it is not difficult \cite{semi3,semi4} to prove
that conditions (\ref{cond1}) are preserved by the flow.

The usefulness of our wave packets stems from the following important
property \cite{raise}.
If we decompose the potential as
$$
E(X)\,=\,W_{a}(X)\,+\,Z_a(X)\,\equiv\,W_{a}(X)\,+\,(E(X)-W_a(X)),
$$
where $W_a(X)$ denotes the second order Taylor
approximation (with the obvious abuse of notation)
$$
W_a(X)\,\equiv\,E(a)\,+\,E^{(1)}(a)\,(X-a)\,+\,E^{(2)}(a)\,(X-a)^2/2
$$
then for all multi-indices $j$,
\bea
&&i\,\hbar\,\frac{\partial }{\partial t}\,
\left[\,\E^{iS(t)/\hbar}\ \ffi_j(A(t),B(t),\hbar,a(t),\eta(t),X)\,\right]
\nonumber\\[8pt] \nonumber
&=&\left(\,-\,\frac{\hbar^2}{2}\,\Delta\,+\,W_{a(t)}(X)\,\right)\,
\left[\,\E^{iS(t)/\hbar}\ \ffi_j(A(t),B(t),\hbar,a(t),\eta(t),X)\,\right] ,
\eea
if $A(t)$, $B(t)$, $a(t)$, $\eta(t)$, and $S(t)$ satisfy (\ref{newton}).
In other words, our semiclassical wave packets $\ffi_j$ exactly
take into account the kinetic
energy and quadratic part $W_{a(t)}(X)$ of the potential when
propagated by means of the classical flow and its linearization around the
classical trajectory selected by the initial conditions.

In the rest of the paper, whenever we write
$\ffi_j(A(t),B(t),\hbar,a(t),\eta(t),X)$, we tacitly assume that
$A(t),B(t),a(t),\eta(t)$, and $S(t)$ are solutions to
(\ref{newton}) with initial conditions satisfying (\ref{cond1}).

\vskip .5cm
\section{The Born--Oppenheimer Expansion in Powers of $\epsilon$}\label{expand}
\setcounter{equation}{0}

\vskip .5cm
In this section we derive an explicit formal expansion in $\eps$
for the solution to the molecular Schr\"odinger equation
by means of a multiple scales analysis. This asymptotic analysis is
similar to that performed, {\it e.g.}, in \cite{masterpiece}.
We discuss this in detail because we need more detailed information on the
structure of successive terms in the expansion.

We start with the molecular Schr\"odinger equation for $d$ nuclear configuration
dimensions,
\be\label{mol}
i\,\eps^2\,\frac{\partial \Psi}{\partial t}\ =\
-\,\frac{\eps^4}2\,\Delta_X\,\Psi\,+\,h(X)\,\Psi.
\ee

We consider the isolated, multiplicity one, smooth eigenvalue $E(X)$ of $h(X)$
of hypothesis {\bf H$_1$}. For the moment we assume $E(X)$ is well defined on
all of $\R^d$ rather than just on a subset $\Xi\subset\R^d$. Later we
introduce a cut-off function to take care of the general case.
We consider the solution
$a(t)$, $\eta(t)$, $A(t)$, $B(t)$, and $S(t)$
to the system (\ref{newton}) of ODE's.
Then, we choose the phase of the
eigenfuction $\widetilde{\Phi}(X,t)$ so that
\be\label{phasechoice}\langle\,\widetilde{\Phi}(X,t),\
(\,i\,\frac{\partial\phantom{t}}{\partial t}\,+\,
i\,\eta(t)\,\nabla_X\,)\,\widetilde{\Phi}(X,t)\,\rangle_{{\cal H}_{el}}\ =\ 0.
\ee
This can always be done. See, {\it e.g.}, \cite{masterpiece}.

The multiple scales analysis consists of separating the two length scales that
are important in the nuclear variable $X$. The electron wave function
is sensitive on an $O(1)$ scale in this variable, so
$X$, or equivalently, $X-a(t)$ is relevant.
The quantum mechanical fluctuations of nuclear wave function occur on an
$O(\eps)$ length scale, so $(X-a(t))/\eps$ is also relevant.
We replace the variable $X$ by both
$w\,=\,X-a(t)$ and $y\,=\,w/\eps$, and consider them as independent
variables. This leads to the new problem of studying
\bea\nonumber
i\,\eps^2\,\frac{\partial\widehat{\Psi}}{\partial t}&=&
\Bigg[\,-\,\frac{\eps^4}2\,\Delta_w\,-\,
\eps^3\,\nabla_w\cdot\nabla_y\,-\,
\frac{\eps^2}2\,\Delta_y\,+\,
i\,\eps^2\,\eta(t)\cdot\nabla_w\,+\,i\,\eps\,\eta(t)\cdot\nabla_y\\
&&\qquad\qquad+\quad\left[\,h(a(t)+w)\,-\,E(a(t)+w)\,\right]\,+\,
E(a(t)+\eps y)\,\Bigg]\
\widehat{\Psi}.\label{multiscale}
\eea
We easily check that if $\widehat{\Psi}(w, y, t)$ solves
(\ref{multiscale}) then $\widehat{\Psi}(X-a(t), (X-a(t))/\eps,t)$ solves
(\ref{mol}).

We define $\Phi(w,t)\,=\,\widetilde{\Phi}(X,t)$. Then (\ref{phasechoice})
becomes
\be\label{wphase}
\langle\,\Phi(w,t),\
i\,\frac{\partial\phantom{t}}{\partial t}\,\Phi(w,t)\,\rangle_{{\cal H}_{el}}
\ =\ 0.
\ee

We seek solutions to (\ref{multiscale}) of the form
$$\widehat{\Psi}(w,y,t)\ =\
e^{i\,S(t)/\eps^2}\,e^{i\,\eta(t)\cdot y/\eps}\,\phi(w,y,t).$$
This requires $\phi(w,y,t)$ to satisfy
\bea\nonumber
i\,\eps^2\,\frac{\partial\phi}{\partial t}&=&
\Bigg[\,-\,\frac{\eps^4}2\,\Delta_w\,-\,
\eps^3\,\nabla_w\cdot\nabla_y\,+\,
\left(\,-\,\frac{\eps^2}2\,\Delta_y\,+\,
  \frac{\eps^2}2\,E^{(2)}(a(t))\,y^2\,\right)\\[5pt] \label{phieqn}
&&\quad+\quad\left[\,h(a(t)+w)\,-\,E(a(t)+w)\,\right]\\[7pt] \nonumber
&&\quad+\quad\left(\,E(a(t)+\eps y)\,-\,E(a(t))-\eps E^{(1)}(a(t))\cdot
y-\eps^2\,E^{(2)}(a(t))\,\frac{y^2}{2!}\,\right)\,\Bigg]\ \phi,
\eea
where here and below we make use of the shorthand notation
$$
E^{(m)}(x)\,\frac{y^m}{m!}\ =\
\sum_{\{k\ :\ |k|=m\}}\,\frac{(D^kE)(x)\,y^k}{k!},
$$
in the usual multi-index notation.
We next assume that $\phi(w,y,t)$ has an expansion of the form
$$\phi(w,y,t)\ =\ \phi_0(w,y,t)\,+\,\eps\,\phi_1(w,y,t)\,+\,
\eps^2\,\phi_2(w,y,t)\,+\,\cdots$$
We further decompose each $\phi_n$ as
$$\phi_n(w,y,t)\ =\ g_n(w,y,t)\,\Phi(w,t)\,+\,\phi_n^{\perp}(w,y,t),$$
by projecting into the $\Phi(w,t)$ direction and into the
orthogonal directions in ${\cal H}_{el}$.

We substitute this expansion into (\ref{phieqn}) and equate terms of the
corresponding powers of $\eps$.

\vskip .3cm
\noindent {\bf Order 0}.\quad The zeroth order terms require
$$\left[\,h(a(t)+w)\,-\,E(a(t)+w)\,\right]\,\phi_0(w,y,t)\ =\ 0.$$
This forces
$$
\phi_0^{\perp}(w,y,t)\ =\ 0.
$$

\vskip .3cm
\noindent {\bf Order 1}.\quad The first order terms require
$$\left[\,h(a(t)+w)\,-\,E(a(t)+w)\,\right]\,\phi_1(w,y,t)\ =\ 0.$$
This forces
$$
\phi_1^{\perp}(w,y,t)\ =\ 0.
$$

\vskip .3cm
\noindent {\bf Order 2}.\quad The second order terms require
$$i\,\frac{\partial\phi_0}{\partial t}\ =\
\left(\,-\,\frac 12\,\Delta_y\,+\,
  E^{(2)}(a(t))\,\frac{y^2}{2!}\,\right)\,\phi_0\,+\,
\left[\,h(a(t)+w)\,-\,E(a(t)+w)\,\right]\,\phi_2.$$
We separately examine the components of this equation in the $\Phi$ direction
and in the orthogonal directions. By (\ref{wphase}), this yields the two
conditions
\be\label{2aeqn}
i\,\frac{\partial g_0}{\partial t}\ =\
\left(\,-\,\frac 12\,\Delta_y\,+\,
  E^{(2)}(a(t))\,\frac{y^2}{2!}\,\right)\,g_0,
\ee
and
\be\label{2beqn}
\left[\,h(a(t)+w)\,-\,E(a(t)+w)\,\right]\,\phi_2\ =\
i\,g_0\,\frac{\partial \Phi}{\partial t}.
\ee
We arbitrarily choose $g_0$ to be the following $w$--independent particular
solution of (\ref{2aeqn}):
\be\label{2aorder}
g_0(w,y,t)\ =\
\eps^{-d/2}\ \sum_{|j|\le J}\ c_{0,j}\ \ffi_j(A(t),B(t),1,0,0,y),
\ee
where $c_{0,j}=c_j$ is determined by the initial conditions.

We let the Hilbert space ${\cal H}_{\mbox{\scriptsize el}}^{\perp}$
be the subspace of ${\cal H}_{\mbox{\scriptsize el}}$
orthogonal to $\Phi(w,t)$. The restriction of
$\left[\,h(a(t)+w)\,-\,E(a(t)+w)\,\right]$
to ${\cal H}_{\mbox{\scriptsize el}}^{\perp}$ is invertible, and we denote
the inverse by
$r(w,t)\ =\ \left[\,h(a(t)+w)\,-\,E(a(t)+w)\,\right]_r^{-1}$.
With this notation, equation (\ref{2beqn}) forces
\bea\nonumber
\phi_2^{\perp}(w,y,t)&=&i\,g_0(w,y,t)\
r(w,t)\ \frac{\partial\Phi}{\partial t}(w,t)\\[5pt]
&=&\eps^{-d/2}\ \sum_{|j|\le J}\ d_{2,j}(w,t)\ \ffi_j(A(t),B(t),1,0,0,y),
\label{2border}
\eea
where
\be\label{d2formula}
d_{2,j}(w,t)\,=\,c_{0,j}\
r(w,t)\ \frac{\partial\Phi}{\partial t}(w,t)
\ee
is ${\cal H}_{\mbox{\scriptsize el}}$--valued.

\vskip .3cm
\noindent {\bf Order 3}.\quad The third order terms require
\bea\nonumber
i\,\frac{\partial\phi_1}{\partial t}&=&
\left(\,-\,\frac 12\,\Delta_y\,+\,
E^{(2)}(a(t))\,\frac{y^2}{2!}\,\right)\,\phi_1\\[5pt] \nonumber
&&\quad-\,
\nabla_w\cdot\nabla_y\,\phi_0\,+\,E^{(3)}(a(t))\frac{y^3}{3!}\,\phi_0\,+\,
\left[\,h(a(t)+w)\,-\,E(a(t)+w)\,\right]\,\phi_3.
\eea
We separately examine the components of this equation in the $\Phi$ direction
and in the orthogonal directions. By (\ref{wphase}), this yields the two
conditions
\bea\nonumber
&&i\,\frac{\partial g_1}{\partial t}\,-\,
\left(\,-\,\frac 12\,\Delta_y\,+\,
  E^{(2)}(a(t))\,\frac{y^2}{2!}\,\right)\,g_1\\
&&\qquad\qquad\qquad\qquad =\ \,
-\ (\nabla_yg_0)\cdot\langle\,\Phi,\,\nabla_w\Phi\,\rangle\,+\,
E^{(3)}(a(t))\,\frac{y^3}{3!}\ g_0,\label{3aeqn}
\eea
and
\be\label{3beqn}
\left[\,h(a(t)+w)\,-\,E(a(t)+w)\,\right]\,\phi_3\ =\
i\,g_1\,\frac{\partial \Phi}{\partial t}\,+\,
(\nabla_yg_0)\,\cdot\,(P_\perp\,\nabla_w\Phi),
\ee
where $P_{\perp}(w,t)$ is the projection in ${\cal H}_{el}$ onto
${\cal H}_{\mbox{\scriptsize el}}^{\perp}$.

The solution to (\ref{3aeqn}) with $g_1(w,y,0)=0$ can be written as
$$
g_1(w,y,t)\ =\
\eps^{-d/2}\ \sum_{|j|\le J+3}\ c_{1,j}(w,t)\ \ffi_j(A(t),B(t),1,0,0,y),
$$
for some coefficients $c_{1,j}(w,t)$.

Equation (\ref{3beqn}) determines
\bea\nonumber
\phi_3^{\perp}(w,y,t)&=&
r(w,t)\,\left(
i\,g_1(w,y,t)\,\frac{\partial\Phi}{\partial t}(w,t)+
(\nabla_y g_0)(w,y,t)\cdot(P_\perp(w,t)
\nabla_w\Phi(w,t))\right)\\[5pt]
&=&\eps^{-d/2}\ \sum_{|j|\le J+3}\ d_{3,j}(w,t)\
\ffi_j(A(t),B(t),1,0,0,y),\nonumber  
\eea
where
\bea\nonumber 
d_{3,j}(w,t)&=&i\,\left(\,r(w,t)\,
\dot{\Phi}(w,t)\,\right)\ c_{1,j}(w,t)\\[5pt]\nonumber
&&+\ \sum_{|q|\le J}\ r(w,t)\ (P_\perp\,\nabla_w\Phi)(w,t)\cdot
\langle\ffi_j,\,\nabla_y\ffi_q\rangle\ c_{0,q}(w,t).
\eea
Here and below $\ds\,\,\,\dot{ }\,\,\equiv \frac{\partial}{\partial t}$.

\vskip .3cm
\noindent {\bf Order ${\bf n}$}.\quad
The $n^{\mbox{\scriptsize th}}$ order terms require
\bea\nonumber i\,\frac{\partial\phi_{n-2}}{\partial t}
&=&\left(\,-\,\frac 12\,\Delta_y\,+\,
  E^{(2)}(a(t))\,\frac{y^2}{2!}\,\right)\,\phi_{n-2}
\,-\,\frac 12\,\Delta_w\,\phi_{n-4}\,-\,
\nabla_w\cdot\nabla_y\,\phi_{n-3}\\[6pt]\nonumber
&&\qquad+\quad
\sum_{m=3}^n\,E^{(m)}(a(t))\,\frac{y^m}{m!}\,\phi_{n-m}\,+\,
\left[\,h(a(t)+w)\,-\,E(a(t)+w)\,\right]\,\phi_n.
\eea
The components of this equation in the $\Phi(w,t)$ direction require
\bea\nonumber
&&i\,\frac{\partial g_{n-2}}{\partial t}\,-\,
\left(\,-\,\frac 12\,\Delta_y\,+\,
  \frac 1{2!}\,E^{(2)}(a(t))\,y^2\,\right)\,g_{n-2}\\[6pt]\nonumber
&&\qquad=\quad-\ \frac 12\,\Delta_w\,g_{n-4}\ -\
\langle\,\Phi,\,\nabla_w\Phi\,\rangle\cdot(\nabla_wg_{n-4})\ -\
\frac 12\,\langle\,\Phi,\,\Delta_w\Phi\,\rangle\,g_{n-4}\\[5pt]\nonumber
&&\qquad\qquad-\ \nabla_w\cdot\nabla_yg_{n-3}\ -\
\langle\,\Phi,\,\nabla_w\Phi\,\rangle\cdot(\nabla_yg_{n-3})\ +\
\sum_{m=3}^n\,E^{(m)}(a(t))\,\frac{y^m}{m!}\ g_{n-m}\\[5pt]
&&\qquad\qquad-\ \frac 12\,\langle\,\Phi,\,\Delta_w\phi_{n-4}^{\perp}\,\rangle
\ -\ \langle\,\Phi,\,\nabla_w\cdot\nabla_y\phi_{n-3}^{\perp}\,\rangle\ +\
i\,\langle\,\frac{\partial\Phi}{\partial t},\,\phi_{n-2}^{\perp}\,\rangle.
\label{naeqn}
\eea
Note that the last term has been transformed from
$\ds -\,i\,\langle\,\Phi,\,
\frac{\partial\phi_{n-2}^{\perp}}{\partial t}\,\rangle$
to
$\ds i\,\langle\,
\frac{\partial\Phi}{\partial t},\,\phi_{n-2}^{\perp}\,\rangle$.
The equivalence of these expressions follows from differentiation of
$\langle\,\Phi,\,\phi_{n-2}^{\perp}\,\rangle\,=\,0$
with respect to $t$.\newline
The components orthogonal to $\Phi(w,t)$ require
\bea\nonumber
&&\left[\,h(a(t)+w)\,-\,E(a(t)+w)\,\right]\,\phi_n\\[5pt]\nonumber
&&=\ P_{\perp}\,\left(\,i\,\frac{\partial \phi_{n-2}^{\perp}}{\partial t}\,\right)\,+\,\left(\,
\frac 12\,\Delta_y\,-\,\frac 1{2!}\,E^{(2)}(a(t))\,y^2\,\right)\,
\phi_{n-2}^{\perp}\,-\,
\sum_{m=3}^n\,E^{(m)}(a(t))\,\frac{y^m}{m!}\,\phi_{n-m}^{\perp}\\[5pt]\nonumber
&&\qquad+\quad\frac 12\,P_{\perp}\,\Delta_w\phi_{n-4}^{\perp}\ +\
(P_{\perp}\,\nabla_w\Phi)\cdot(\nabla_wg_{n-4})\ +\
\frac 12\,(P_{\perp}\,\Delta_w\Phi)\,g_{n-4}\\[5pt]
&&\qquad+\quad P_{\perp}\,\nabla_w\cdot\nabla_y\phi_{n-3}^{\perp}\ +\
(P_{\perp}\,\nabla_w\Phi)\cdot(\nabla_yg_{n-3})\ +\
i\,\frac{\partial\Phi}{\partial t}\,g_{n-2}.
\label{nbeqn}
\eea

Equation (\ref{nbeqn}) determines $\phi_{n}^{\perp}(w,y,t)$ by an
application of $\left[h(a(t)+w)-E(a(t)+w)\right]_r^{-1}\!$.

It is easily checked that the solution to (\ref{naeqn}) with
$g_{n-2}(w,y,0)\,=\,0$ has the form
\be\label{naorder}
g_{n-2}(w,y,t)\ =\
\eps^{-d/2}\ \sum_{|j|\le J+3n-6}\ c_{n-2,j}(w,t)\ \ffi_j(A(t),B(t),1,0,0,y),
\ee
for some coefficients $c_{n-2,j}(w,t)$, and that the $y$
dependence of the vector $\phi_{n}^{\perp}$ has the same form,
with other coefficients depending on $(w,t)$, {\it i.e.},
\be\label{nborder}
\phi_n^{\perp}(w,y,t)\ =\
\eps^{-d/2}\ \sum_{|j|\le J+3n-6}\ d_{n,j}(w,t)\
\ffi_j(A(t),B(t),1,0,0,y),
\ee
where the $d_{n,j}(w,t)$ take their values in the electronic Hilbert
space.

Equations (\ref{naeqn}) and (\ref{nbeqn}) determine $c_{n-2,j}$ and
$d_{n,j}$. When recursively solving these equations, we must
determine $d_{n,j}$ before $c_{n,j}$ because the right hand side of
(\ref{naeqn}) (with $n-2$ replaced by $n$) contains $\phi_n^\perp$.

The solution to (\ref{nbeqn}) in terms of the $d_{n,j}$, is
\be\label{ddecomp}
d_{n,j}(w,t)\ =\ \sum_{i=1}^8\ \Delta_i(w,t),
\ee
where
\bea\nonumber
\Delta_1(w,t)&=&i\,r(w,t)\,P_\perp(w,t)\ \dot{d}_{n-2,j}(w,t)
\\[7pt]\nonumber
\Delta_2(w,t)&=&-\ \sum_{3\le|m|\le n}\frac{(D^mE)(a(t))}{m!}
\sum_{|q|\le J+3(n-|m|-2)}\langle\ffi_j,\,y^m\ffi_q\rangle\
r(w,t)\ d_{n-|m|,q}(w,t)
\\[7pt]\nonumber
\Delta_3(w,t)&=&\frac 12\,r(w,t)\,P_\perp(w,t)\ (\Delta_wd_{n-4,j})(w,t)
\\[7pt]\nonumber
\Delta_4(w,t)&=&r(w,t)\,P_\perp(w,t)\
(\nabla_w\Phi)\cdot(\nabla_wc_{n-4,j})(w,t)
\\[7pt]\nonumber
\Delta_5(w,t)&=&\frac 12\,r(w,t)\,P_\perp(w,t)\ (\Delta_w\Phi)\
c_{n-4,j}(w,t)
\\[7pt]\nonumber
\Delta_6(w,t)&=&\sum_{|q|\le J+3(n-5)}\,r(w,t)\,P_\perp(w,t)\
\langle\ffi_j,\,\nabla_y\ffi_q\rangle\ (\nabla_wd_{n-3,q})(w,t)
\\[7pt]\nonumber
\Delta_7(w,t)&=&\sum_{|q|\le J+3(n-3)}\,r(w,t)\,P_\perp(w,t)\
(\nabla_w\Phi)\ \langle\ffi_j,\,\nabla_y\ffi_q\rangle\
c_{n-3,q}(w,t)
\\[7pt]\nonumber
\Delta_8(w,t)&=&i\,r(w,t)\,P_\perp(w,t)\ \dot{\Phi}(w,t)\ c_{n-2,j}(w,t).
\eea
Similarly, the solution to (\ref{naeqn}) in terms of the $c_{n,j}$ is
obtained by integration with respect to $t$ of $i\,\dot{c}_{n,j}(w,t)$,
where
\be\label{cdecomp}
i\,\dot{c}_{n,j}(w,t)\ =\ \sum_{i=1}^9\ \Gamma_i(w,t),
\ee
where
\bea\nonumber
\Gamma_1(w,t)&=&-\ \frac 12\ (\Delta_wc_{n-2,j})(w,t)
\\[7pt]\nonumber
\Gamma_2(w,t)&=&-\ \langle\Phi,\,\nabla_w\Phi\rangle\cdot
(\nabla_wc_{n-2,j})(w,t)
\\[7pt]\nonumber
\Gamma_3(w,t)&=&-\ \frac 12\ \langle\Phi,\,\Delta_w\Phi\rangle\
c_{n-2,j}(w,t)
\\[7pt]\nonumber
\Gamma_4(w,t)&=&-\ \sum_{|q|\le J+3(n-1)}\,
\langle\ffi_j,\,\nabla_y\ffi_q\rangle\cdot(\nabla_wc_{n-1,q})(w,t)
\\[7pt]\nonumber
\Gamma_5(w,t)&=&-\ \sum_{|q|\le J+3(n-1)}\,
\langle\Phi,\,\nabla_w\Phi\rangle\cdot\langle\ffi_j,\,\nabla_y\ffi_q\rangle\
c_{n-1,q}(w,t)
\\[7pt]\nonumber
\Gamma_6(w,t)&=&\sum_{3\le|m|\le n+2}\ \sum_{|q|\le J+3(n+2-m)}\,
\frac{(D^mE)(a(t))}{m!}\ \langle\ffi_j,\,y^m\ffi_q\rangle\
c_{n+2-m,q}(w,t)
\\[7pt]\nonumber
\Gamma_7(w,t)&=&\frac 12\
\langle\Phi,\,(\Delta_wd_{n-2,j})(w,t)\rangle
\\[7pt]\nonumber
\Gamma_8(w,t)&=&-\ \sum_{|q|\le J+3(n-3)}\
\langle\ffi_j,\,\nabla_y\ffi_q\rangle\cdot
\langle\Phi,\,(\nabla_wd_{n-2,q})(w,t)\rangle
\\[7pt]\nonumber
\Gamma_9(w,t)&=&i\ \langle\dot{\Phi},\,d_{n,j}(w,t)\rangle .
\eea

\vskip .5cm
\section{The Main Result}\label{mr}
\setcounter{equation}{0}

We introduce a $C^{\infty}$ real valued cut-off function
$F:\R^d\ra\R$ that equals $1$ in a neighborhood
of the origin and equals zero away from the origin. More precisely,
we choose
$0<b_0<b_1<\infty$, such that
$$
\mbox{supp}\ (\partial_{w_i}F)(w)\,\subseteq\,\{w\in\R^d\;:\; b_0<|w|<b_1\},
$$
for any $i\in\{1,\dots,d\}$, and such that for any $t\in\Omega$,
all quantities appearing in the above expansion are well defined for
$w\in\R^d$ with $|w|<b_1$.
Here $\Omega$ is a particular simply connected open complex neighborhood of the
real interval $[0,\,T]$ that we construct in Section \ref{AP}
under hypotheses {\bf H}$_0$ and {\bf H}$_1$.

We define our approximate solution to (\ref{mol}) at order $N$
by the following expression:
\bea\label{approx}
&&\hat{\Psi}_N(w, y, t)\\[5pt]
&=&F(w)\,\E^{iS(t)/\eps^2}\,\E^{i\eta(t)\cdot y/\eps}\,
\left(\,\sum_{n=0}^N\,\eps^n\,g_n(w,y,t)\,\Phi(w,t)\ +\
\sum_{n=2}^{N+2}\,\eps^n\,\phi_n^\perp(w,y,t)\right).\nonumber
\eea
We prove in Section \ref{optimal} that
this quantity agrees with an exact solution up to an error whose norm is bounded
by $\eps^N$ for $t\in [0,\,T]$.

We emphasize that once the molecular
hamiltonian $h(X)$ and its spectral data $E(X)$, $\Phi(X)$ are given,
the only arbitrary input of the above derived expansion consists
of the set of coefficients $c_{0,j}$, $|j|\leq J$.
We note that at time $t=0$, we have
$c_{n,j}(0,w)\equiv 0$ for all $n\geq 1$. Thus, at $t=0$, the approximation
reduces to
\bea\nonumber 
&&\hat{\Psi}_N(w, y, 0)\\
&=&F(w)\,\E^{iS(0)/\eps^2}\,\E^{i\eta(0)\cdot y/\eps}\,
\left(\,g_0(0,y,0)\,\Phi(w, 0)\,+\,\sum_{n=2}^{N+2}\,\eps^n\,
\phi_n^\perp(w,y,0)\,\right).\nonumber
\eea
This expression is completely determined by $g_0(0,y,0)$, the nuclear
part of the wave function parallel to the chosen electronic level at time $0$.

As is usual in the study of adiabatic problems, in order to get accurate
information on the evolution of an initial wave function associated with
a specific electronic level, one needs to include a higher order
component perpendicular to that electronic level.
This higher order part is completely determined by the parallel
part. Here it is given (up to phase and cut-off functions) at time $0$ by
$\ds\sum_{n=2}^{N+2}\,\eps^n\,\phi_n^\perp(w, y, 0)$.

\vskip .4cm
We now state our main Theorem:

\begin{thm}\label{main}
Assume hypotheses {\bf H}$_0$ and {\bf H}$_1$ and consider the above
construction. For all sufficiently small choices of
$g>0$, there exist $C(g)>0$ and $\Gamma(g)>0$ such that,
for $N(\eps)=\grintl g^2/\eps^2\grintr$,
the vector $\Psi_*(X,t,\eps)\,=\,\hat{\Psi}_{N(\eps)}(X-a(t),(X-a(t))/\eps,t)$
satisfies
$$
\left\|\,\E^{-itH(\eps)/\eps^2}\,\Psi_*(X,0,\eps)\,-\,\Psi_*(X,t,\eps)\,
\right\|_{L^2({\smallR}^d,{\cal H}_{\mbox{\scriptsize el}})}
\ \leq\ C(g)\ \E^{-\Gamma(g)/\eps^2},
$$
for all $t\in [0,\,T]$, as $\eps\ra 0$.\\
Moreover, we have the following exponential localization result.
For any $b>0$ and a sufficiently small choice of $g>0$ (that depends on $b$),
there exist $c(g)$ and\, $\gamma(g)>0$, such that
$$
\left(\,\int_{|x-a(t)|>b}\,
\|\Psi_*(X,t,\eps)\|^2_{{\cal H}_{\mbox{\scriptsize el}}}\ dx\,\right)^{1/2}
\ \leq\ c(g)\ \E^{-\gamma(g)/\eps^2},
$$
for all $t\in [0,\,T]$, as $\eps\ra 0$.
\end{thm}

\vspace{.2cm}
The strategy of the proof is as follows:\ We
consider the approximation $\hat{\Psi}_N(X-a(t),(X-a(t))/\eps,t)$
and the exact solution to the Schr\"odinger equation with the same initial
conditions.
We estimate the norm of the error (that is the difference between these
two quantities)
as a function of both $N$ and $\eps$. Apart from some
subtleties, the norm of the error is bounded by
$C\eps^N (\tau N^{1/2})^N$, for some constants $C$ and $\tau>0$. We minimize the
error estimate over all choices of $N$. This yields $N\simeq g^2/\eps^2$, for
sufficiently small $g>0$, and an estimate of order $\E^{-\Gamma(g)/\eps^2}$ for
the norm of the error.

\vspace{.4cm}
We prove two extensions of this result in Section \ref{gene}.
In the first extension, we consider the validity of our approximation
on the Ehrenfest time scale, {\it i.e.}, when $T=T(\eps)\simeq \ln(1/\eps)$.
In the second extension, we study the dependence of our construction on $J$,
in order to extend our main result to a wider class of initial conditions.
We refer the reader to Section \ref{gene} for the precise statements.

\vskip .5cm
\section{Analyticity Properties}\label{AP}
\setcounter{equation}{0}

Our estimates depend on analyticity in $t\in\Omega$ of the vectors
$c_{n}(w,t)\in l^2(\N^d,\C)$ and
$d_{n}(w,t)\in l^2(\N^d,{\cal H}_{\mbox{\scriptsize el}})$, where
$\Omega$ is the particular simply connected open complex neighborhood of the
real interval $[0,\,T]$ mentioned at the beginning of Section \ref{mr}.

To construct $\Omega$, we begin with several observations.
Our hypotheses imply that the eigenvalue $E(X)$ is analytic in
$\Sigma_{\delta'}$, so the solutions $a(t)$, $\eta(t)$, $A(t)$, $B(t)$, and
$S(t)$ are well defined for all $t\in [0,\,T]$. Moreover, by standard
arguments \cite{dieudonne}, these functions all have analytic
continuations from $[0,\,T]$ to a simply connected open set $\Omega_1$ that
contains $[0,\,T]$.
We assume without loss of generality that $\Omega_1=\overline{\Omega_1}$, where
$\overline{\Omega_1}$ denotes the conjugate of $\Omega_1$.

We note that $A^*(t)$ and $B^*(t)$ also have analytic continuations from
$[0,\,T]$ to $\Omega_1$.
To see this for $A^*(t)$, note that for $t\in[0,\,T]$,
$A^*(t)\,=\,A^*(\overline{t})$, and $A^*(\overline{t})$ has an analytic
continutation to $\overline{\Omega_1}$. The argument for $B^*(t)$ is
similar.

It now follows easily from the definitions that for each $X$,
$\ffi_j(A(t),B(t),\eps^2,a(t),\eta(t),X)$ and
$\overline{\ffi_j(A(t),B(t),\eps^2,a(t),\eta(t),X)}$ have analytic
continuations from $[0,\,T]$ to some simply connected open set $\Omega_2$. For
$t\in[0,\,T]$, the real
part of $B(t)A(t)^{-1}$ is strictly positive. This positivity will remain true
for the real part of the analytic continuation of $B(t)A(t)^{-1}$ on some
simply connected subset $\Omega\subset\Omega_1\cap\Omega_2$ that contains
$[0,\,T]$. We assume without loss of generality that
$\Omega=\overline{\Omega}$ and we can assume that $\Omega$ has the form
$\{\,t\,:\,-a<\mbox{Re}\,t<b\mbox{ and }|\mbox{Im}\,t|<c\,\}$ where
$a>c>0$ and $b>T+c$.
It follows that for $t\in\Omega$, both
$\ffi_j(A(t),B(t),\eps^2,a(t),\eta(t),x)$ and
$\overline{\ffi_j(A(t),B(t),\eps^2,a(t),\eta(t),x)}$ have analytic
continuations from $[0,\,T]$ to $\Omega$ as elements of $L^2(\R^d)$.

Using these results and carefully
examining the constructions of the vectors $c_n(w,t)$ and $d_n(w,t)$,
we see that they are analytic in $t$ for $t\in\Omega$, also. \\

Our hypotheses on $h(\cdot)$ and the above results also
show that each of the following quantities is analytic in $t$ for
$t\in \Omega$ and each fixed $w\in\Sigma_\delta\subset\C^d$, for
sufficiently small $\delta$:
\bea\nonumber
&&r(w,t)\ =\ \left[\,h(a(t)+w)\,-\,E(a(t)+w)\,\right]_r^{-1},\\[4pt] \nonumber
&&{\Phi}(w,t),\\[4pt] \nonumber
&&(D_w^\alpha{\Phi})(w,t),\quad\mbox{for } |\alpha|\le 2,\\[4pt]
\nonumber
&&D_w^\alpha E(a(t)), \quad\mbox{for all}\quad\alpha\\[4pt] \nonumber
&&P_\perp(w,t).
\eea
By explicit computation of the phase corresponding to (\ref{phasechoice}) 
it is easy to check that $\Phi(w,t)$ and its derivatives are also analytic
for $t\in\Omega$.

Moreover, if $f_i(w,t)$, ($i$ in some finite set)
represents any of these quantities, $f_i$ is analytic in $w\in
\Sigma_\delta$, for any fixed $t\in \Omega$.
Thus, by the Cauchy integral formula, we can assume that the following bounds
hold (with the appropriate norm in each case):
\be\label{bound}
\left\|\,(D^\alpha_wf_i)(w,t)\,\right\|\ \le\
c_i\ G_i^{|\alpha|}\ \frac{\alpha !}{(1+|\alpha|)^{d+1}},
\ee
for some $c_i$, $G_i$,
$w\in\Sigma_\delta$, and $\alpha$ ranges over all multi-indices.

We can assume here that all $G_{i}\leq D_{2}$ for some
constant
$$
D_{2}\ \geq\ 1,
$$
and we associate the prefactors $c_{i}$ in (\ref{bound}) with the different
functions according to the rules
$$\matrix{
c_{1} & \leftrightarrow & rP_{\perp}&\qquad\qquad&c_{2} & \leftrightarrow &
\Phi\cr\vspace{4pt}
c_{3} & \leftrightarrow & \dot{\Phi}&&c_{4} & \leftrightarrow &
\nabla_{w}\Phi\cr\vspace{4pt}
c_{5} & \leftrightarrow & \Delta_{w}\Phi&&c_{6}&\leftrightarrow & E\cr
\vspace{4pt}
c_{4}' & \leftrightarrow & \bra \Phi,\,\nabla_{w}\Phi\ket&\qquad&
c_{5}' & \leftrightarrow & \bra \Phi,\,\Delta_{w}\Phi\ket .}
$$

\vskip .5cm
\section{Structure and Estimates of the $c_{n}(w,t)$ and $d_n(w,t)$}
\setcounter{equation}{0}

In this section, we decompose the functions $g_n$ and
$\phi_n^{\perp}$ of Section \ref{expand} into pieces, each of which satisfies
various estimates.

Throughout this section, all $w$--dependent quantities
are defined for $w$ in the support of the cut-off function $F$.
Furthermore, all the results of this section are claimed to hold only on
the support of $F$.

Our decompositions of $g_n(w,y,t)$ and $\phi_n^{\perp}(w,y,t)$ have the
following forms:
\bea\label{messg}
\!\!\!\!\!&&g_n(w,y,t)\\[5pt] \nonumber
\!\!\!\!\!&=&\eps^{-d/2}\
\sum_{\beta\in {\cal B}_{n,1}}\
\sum_{p\le n}\
\sum_{|l|+k\le p+\frac n2}\
\sum_{|j|\le J+n+2(p-|l|-k)}\
c_{n,p,l,k,\beta,j}(w,t)\ \ffi_j(A(t),B(t),1,0,0,y).
\eea
and
\bea\label{messperp}
&&\!\!\!\!\!\!\!\!\!\!\!\!\phi_n^\perp(w,y,t)\ =\\[5pt] \nonumber
&&\!\!\!\!\!\!\!\!\!\!\!\!\!\!\eps^{-d/2}
\sum_{\beta\in {\cal B}_{n,2}}\,
\sum_{p\le n-1}\,
\sum_{|l|+k\le p+\frac{n-1}2}\
\sum_{|j|\le J+(n-1)+2(p-|l|-k)}\
d_{n,p,l,k,\beta,j}(w,t)\ \ffi_j(A(t),B(t),1,0,0,y).
\eea
In (\ref{messg}), $n$, $k$ and $p$ are non-negative integers; $j$ and $l$ are
multi-indices; and the index $\beta$ runs over a finite set
${\cal B}_{n,1}$.
The number $J$ is fixed by the initial conditions.
Each $c_{n,p,l,k,\beta,j}$ is a complex valued function.\\
In (\ref{messperp}), $n\ge 2$, $k$ and $p$ are non-negative integers; $j$ and
$l$ are multi-indices; and the index $\beta$ runs over a finite set
${\cal B}_{n,2}$.
Each $d_{n,p,l,k,\beta,j}(w,t)$ takes values in ${\cal H}_{el}$.

We let $c_{n,p,l,k,\beta}(w,t)$ and $d_{n,p,l,k,\beta}(w,t)$
respectively denote vectors in $l^2(\N^d,\C)$
and\\ $l^2(\N^d,{\cal H}_{el})$
whose components are $c_{n,p,l,k,\beta,j}(w,t)$ and
$d_{n,p,l,k,\beta,j}(w,t)$.

\vskip .3cm
The crucial step in the proof of Theorem \ref{main} is the following:

\begin{prop}\label{mainprop}
There is a recursive construction of the coefficients $c_{n,p,l,k,\beta,j}(w,t)$
and $d_{n,p,l,k,\beta,j}(w,t)$ for $w$ on the support of $F$.\\
The indices for $c_{n,p,l,k,\beta,j}(w,t)$ are non-negative and satisfy
\bea\nonumber
\beta&\in&{\cal B}_{n,1}\\[3pt] \nonumber
p&\le&n,\\[3pt] \nonumber
|l|+k&\le&p+\frac{n}2,\\[3pt] \nonumber
|j|&\le&J+n+2(p-|l|-k).
\eea
The indices for $d_{n,p,l,k,\beta,j}(w,t)$ are non-negative and satisfy
\bea\nonumber
n&\ge&2\\[3pt] \nonumber
\beta&\in&{\cal B}_{n,2},\\[3pt] \nonumber
p&\le&n-1\\[3pt] \nonumber
|l|+k&\le&p+\frac{n-1}2\\[3pt] \nonumber
|j|&\le&J+(n-1)+2(p-|l|-k).
\eea
Moreover, the following conditions are satisfied:\\
i)\quad For any $n>0$,\quad $c_{n,0,l,k,\beta,j}(w,t)\,=\,0$.\\
ii)\quad There exists $K_0>0$, such that the number of terms in both of the
sums (\ref{messg}) and (\ref{messperp}) is bounded by $\E^{K_0 n}$.\\
iii)\quad For $t\in\Omega$, let $\dist(t)$ be the distance from $t$ to the
complement of $\Omega$.
The coefficients $c_{n,p,l,k,\beta}(w,t)$ and $d_{n,p,l,k,\beta}(w,t)$
are analytic for $t\in\Omega$, and there exist constants $D_1$ and
$D_2$, such that
\bea\label{cestimate}
&&\|(D_w^\alpha c_{n,p,l,k,\beta})(w,t)\|\\[7pt] \nonumber
&\le&D_1\ D_2^{|\alpha|+|l|+4n}\ \frac{(\alpha+l)!}{(1+|\alpha|)^{d+1}}\
\frac{|t|^p}{p!}\ \frac{k^k}{\dist(t)^k}\
\left[\,\frac{(J+n+2(p-|l|-k))!}{J!}\,\right]^{1/2},
\eea
and
\bea\label{destimate}
&&\!\!\!\!\!\!\!\!\!\!\!\!\|(D_w^\alpha d_{n,p,l,k,\beta})(w,t)\|\\ \nonumber
&&\!\!\!\!\!\!\!\!\!\!\!\!\!\!\!\!\!\le\
D_1D_2^{|\alpha|+|l|+4(n-1)}\,\frac{(\alpha+l)!}{(1+|\alpha|)^{d+1}}\
\frac{|t|^p}{p!}\ \frac{k^k}{\dist(t)^k}\
\left[\frac{(J+(n-1)+2(p-|l|-k))!}{J!}\right]^{1/2}.
\eea
\end{prop}

\vskip .3cm \noindent {\bf Remark}\quad
The complicated estimates (\ref{cestimate}) and (\ref{destimate}) are motivated
by estimates
used in semiclassical approximations and adiabatic approximations. The factors
on the right hand sides that explicitly involve $J$, $n$, and $p$ occur in the
semiclassical paper \cite{hagjoy5}. The factors
that involve $\alpha$ and $l$ appear in the adiabatic paper \cite{nenciu}. The
factors that involve
$k$ occur in a proof of the adiabatic results of \cite{nenciu} that are based on
Cauchy estimates instead of Nenciu's lemma \cite{nenciu} (that we generalize
below as Lemma \ref{nenciuprop2}). We were unable to prove Proposition
\ref{mainprop} without using a combination of all of these techniques.
We estimate adiabatic error terms by using Nenciu's approach in the $w$
variable and Cauchy estimates in the $t$ variable.

\vskip .5cm
\subsection{The Toolbox}\label{tool}

To prove Proposition \ref{mainprop}, we repeatedly use the following
very handy lemmas, whose proofs are given in Section \ref{techie}. The first
two lemmas deal with basic properties of analytic functions of one variable
and are consequences of the Cauchy integral formula.

\vskip .5cm
\begin{lem}\label{cauchylem}
For $k=0$, define $k^k=1$.
Suppose $g$ is an analytic vector--valued function on the strip
$S_\delta\ =\ \{\,t\,:\,|\mbox{\rm Im}\,t|<\delta\,\}$.
If $g$ satisfies
$$\|g(t)\|\ \le\ C\,k^k\,(\delta\,-\,|\mbox{\rm Im}\,t|)^{-k},$$
for some $k\ge 0$, then $g'$ satisfies
$$\|g'(t)\|\ \le\ C\,(k+1)^{k+1}\,(\delta\,-\,|\mbox{\rm
Im}\,t|)^{-k-1},$$
for all $t\in S_\delta$.
\end{lem}

\vskip .5cm
Lemma \ref{cauchylem} has a generalization to regions other than infinite
strips. The generalization is needed if one wishes to study problems
where analyticity holds only in a neighborhood of a finite time interval.
The proof of the generalized lemma is similar to that of Lemma
\ref{cauchylem}, but involves slightly more complicated geometry. The
precise statement is the following:

\vskip .5cm
\begin{lem}\label{upgrade}
For $k=0$, define $k^k=1$.
Suppose $g$ is an analytic vector--valued function in an open region
$\Omega\subset\C$. For $t\in\Omega$, let $\dist(t)$ be the distance
from $t$ to $\Omega^C$, the complement of $\Omega$.
If $g$ satisfies
$$\|g(t)\|\ \le\ C\,k^k\,(\dist(t))^{-k},$$
for all $t\in\Omega$ and some $k\ge 0$, then $g'$ satisfies
$$\|g'(t)\|\ \le\ C\,(k+1)^{k+1}\,(\dist(t))^{-k-1},$$
for all $t\in\Omega$.
\end{lem}

The next lemma gives estimates on indefinite integrals of
certain analytic functions under stronger assumptions on the domain
$\Omega$.

\vskip .5cm
\begin{lem}\label{integrate}
Suppose $f$ is an analytic vector--valued function in an open region
$\Omega\subset\C$. For $t\in\Omega$, let $\dist(t)$ be the distance
from $t$ to $\Omega^C$. We assume the domain is
star-shaped with respect to the origin and that the origin is
the most distant point to $\Omega^C$, {\it i.e.}, $\dist(0)\geq\dist(t)$,
for all $t\in \Omega$. Moreover, we assume that $\dist(t)$ is monotone
decreasing along any line emanating from the origin.
If $f$ satisfies
$$\|f(t)\|\ \le\ C\ |t|^p\ (\dist(t))^{-k},$$
for all $t\in\Omega$ and some $k\ge 0$, then
$\left\|\,\int_0^t\,f(s)\,ds\,\right\|$
satisfies
$$\left\|\,\int_0^t\,f(s)\,ds\,\right\|\ \le\
C\ \frac{|t|^{p+1}}{p+1}\ (\dist(t))^{-k},$$ for all $t\in\Omega$.
\end{lem}

\vskip .3cm \noindent
{\bf Remark:}\quad In our situation, examples of sets $\Omega$ we can use
that satisfy the conditions of Lemma \ref{integrate} are
infinite symmetrical horizontal strips or the rectangular regions chosen in
Section \ref{AP}.

\vskip .5cm
A fourth tool we repeatedly use below is a multidimensional
generalization of a lemma used in \cite{nenciu}.
We warn the reader that the symbol
for a norm means different things in different contexts, {\it e.g.}, for
scalar--valued, operator--valued, and vector--valued functions, it
respectively means absolute value, operator norm, and vector space norm.

\vskip .5cm
\begin{lem}\label{nenciuprop2}
The quantity
\be\label{nudef}
\nu\ =\ \sup_{\alpha}\quad
(1+|\alpha|)^{d+1}\ \sum_{\{\,l\,:\,0\,\le\,l_i\,\le\,\alpha_i\,\}}\
\frac{1}{(1+|l|)^{d+1}}\ \frac{1}{(1+|\alpha-l|)^{d+1}}.
\ee
is finite.\\
Let $\Sigma$ be an open subset of $\C^d$.
Suppose $M(\cdot)\in C^\infty(\Sigma)$ is scalar--valued or
operator--valued, and
$N(\cdot)\in C^\infty(\Sigma)$ is either operator--valued or
vector--valued. Assume these functions satisfy
\bea
\left\|\,\left(D^\alpha\,M\right)(x)\,\right\|&\le&m(x)\ a(x)^{|\alpha+p|}\quad
\frac{(\alpha+p)!}{(1+|\alpha|)^{d+1}}\\[5pt]
\left\|\,\left(D^\alpha\,N\right)(x)\,\right\|&\le&n(x)\ a(x)^{|\alpha+q|}\quad
\frac{(\alpha+q)!}{(1+|\alpha|)^{d+1}}
\eea
for $x\in\Sigma$, all multi-indices $\alpha$, and some fixed multi-indices
$p$ and $q$. Then
\be
\left\|\,\left(\,D^\alpha\,(M\,N)\,\right)(x)\,\right\|\
\le\ m(x)\ n(x)\ \nu\ a(x)^{|\alpha+p+q|}\quad
\frac{(\alpha+p+q)!}{(1+|\alpha|)^{d+1}}
\ee
for each multi-index $\alpha$, where $\nu$ is defined by (\ref{nudef}).
\end{lem}

\vskip .5cm
\subsection{Proof of Proposition \ref{mainprop}}

We prove Proposition \ref{mainprop} by induction and begin with the
case $n=0$. We construct $c_{0,0,0,0,\beta,j}\equiv c_{0,j}$
with $\beta=1\in{\cal B}_{0,1}\equiv\{1\}$.
We note that there is no $d_{n,p,l,k,\beta}(w,t)$ for $n\le 1$; the inequalities
for its indices in the conclusion to the proposition cannot be satisfied by
non-negative integers. Whenever $d_{n,p,l,k,\beta}(w,t)$ with $n\le 1$ appears
in any of the formal calculations below, it is understood to be zero.

We now assume that the estimates (\ref{cestimate}) and (\ref{destimate})
on $c_{m,p,l,k,\beta}(w,t)$ and $d_{m,p,l,k,\beta }(w,t)$ are
true for all $m\leq n-1$ and prove they still hold for $m=n$.

Our strategy is to show that each contribution $\Delta_{i}$
and $\Gamma_{i}$ consists of a finite sum
of terms that satisfy the required estimate. We estimate the
number of terms by a separate argument. Our main tools are
Lemmas \ref{ylem}, \ref{upgrade}, \ref{integrate}, and \ref{nenciuprop2}.

The index $\beta$ must be considered when counting the number of terms, but
it plays no role in the estimates of the individual terms. To simplify the
notation, we drop it while estimating the terms.




\vspace{.5cm}
\noindent {\bf The Term $\Delta_{1}$}\\

We begin by considering the contribution to (\ref{destimate}) from
the term $\Delta_{1}$ in (\ref{ddecomp}).

By induction, each $d_{n-2,p,l,k,\beta}(w,t)$ is analytic for $t\in\Omega$ and
has a $p^{\mbox{\scriptsize th}}$ order zero at $t=0$. It follows that
$d_{n-2,p,l,k,\beta}(w,t)\,=\,t^p\,f(t)$, where $f$ is analytic in $\Omega$.
When we take the time derivative, we obtain two terms,
$p\,t^{p-1}\,f(t)$ and $t^p\,\dot{f}(t)$. These, respectively, give rise to
two terms $d_{n,p-1,l,k,\beta'}(w,t)$ and $d_{n,p,l,k+1,\beta''}(w,t)$.

We consider all $w$-derivatives of $\Delta_{1}(w,t)$. We
apply the induction hypothesis, Lemma \ref{upgrade}, and Lemma \ref{nenciuprop2}
to obtain
\bea\nonumber
&&\!\!\!\!\!\!\!\!\!\!\!\!
\left\|\,D_{w}^{\alpha}\left( rP_{\perp}(w,t)\dot{d}_{n-2, p, l, k}(w,t)
\right)\,\right\|\\[5pt] \nonumber
&&\!\!\!\!\!\!\!\!\!\!\!\!
\leq\ c_{1}\,\nu\,D_{1}\,D_{2}^{|\alpha |+|l|+4(n-3)}\
\frac{(\alpha+l)!}{(1+|\alpha|)^{d+1}}\
\sqrt{\frac{(J+(n-3)+2(p-|l|-k))!}{J!}}\\[5pt] \nonumber
&&\quad\quad\quad\quad\quad\quad\quad\quad\quad\quad\quad\quad\quad\quad
\qquad\times\quad
\left(\,\frac{|t|^{p-1}}{(p-1)!}\ \frac{k^{k}}{\dist(t)^{k}}\ +\
\frac{t^{p}}{p!}\ \frac{(k+1)^{k+1}}{\dist(t)^{k+1}}\,\right)\\[5pt]
\nonumber
&&\!\!\!\!\!\!\!\!\!\!\!\!
=\,c_{1}\nu D_{1}D_{2}^{-8}D_{2}^{|\alpha |+|l|+4(n-1)}\,
\frac{(\alpha+l)!}{(1+|\alpha |)^{d+1}}\,
\frac{|t|^{p'}}{p'!}\,\frac{k^{k}}{\dist(t)^{k}}\,
\sqrt{\frac{(J+(n-1)+2(p'-|l|-k))!}{J!}}\\[5pt] \nonumber
&&\!\!\!\!\!\!\!\!\!\!\!\!
+\,c_{1}\nu D_{1}D_{2}^{-8}D_{2}^{|\alpha |+|l|+4(n-1)}\,
\frac{(\alpha+l)!}{(1+|\alpha |)^{d+1}}\,
\frac{|t|^{p}}{p!}\,\frac{{k'}^{k'}}{\dist(t)^{k'}}\,
\sqrt{\frac{(J+(n-1)+2(p-|l|-k'))!}{J!}},
\eea
with $p'=p-1$ and $k'=k+1$. We check that
\bea\nonumber
p&\leq&n-3\quad <\quad n-1,\\[3pt] \nonumber
p'&\leq&n-4\quad <\quad n-1,\\[3pt] \nonumber
|l|+k&\leq& p+(n-3)/2\quad =\quad p'+(n-1)/2,\\[3pt] \nonumber
|l|+k'&\leq&p+(n-3)/2+1\quad =\quad p+(n-1)/2,
\eea
and the ranges of the components of each vector satisfy
\bea\nonumber
|j|&\leq &J+(n-1)+2(p'-|l|-k),\\[3pt] \nonumber
|j|&\leq &J+(n-1)+2(p-|l|-k'),
\eea
as required.
Hence, we get the desired bound for each of the two contributions, provided
$$
D_{2}^8\ \geq\ c_{1}\ \nu.
$$
\vspace{.5cm}
\noindent {\bf The Term $\Delta_{2}$}\\

In the analysis of this term, we encounter an infinite matrix that represents
multiplication by $y^m$ in the basis of semiclassical wave packets.
We denote this matrix by $\bra\ffi,\,y^{m}\ffi\ket$. Its entries are
$\bra\ffi_j,\,y^{m}\ffi_q\ket(t)$, for multi-indices $m, j, q\in \N^{d}$.
We recall that Lemma \ref{ylem} gives bounds
for these matrix elements and also states that
$\bra\ffi_j,\,y^{m}\ffi_q\ket(t)=0$ if $\left|\,|j|-|q|\,\right|>|m|$.

We adopt the analogous
notation for the infinite matrix $\bra\ffi,\,D_y^m\ffi\ket$ that represents
the operator $D_{y}^{m}$ in the basis of semiclassical wave packets.

We define $d_{0}=\sqrt{2}\,d$.
Then, using (\ref{bound}),
Lemmas \ref{ylem}, \ref{integrate}, \ref{upgrade}, and \ref{nenciuprop2},
and some algebra, we obtain
\bea\nonumber
&&\left\|\,D_{w}^{\alpha}\,\sum_{\tilde{m}=3}^{n}\,\sum_{|m|=\tilde{m}}
\frac{D^{m}E(a(t))}{m!}\,\bra\ffi,\,y^{m}\ffi\ket\,(rP_{\perp})(w,t)\,
d_{n-\tilde{m},p,l,k}(w,t)\,\right\|\\[5pt] \nonumber
&\leq&
\sum_{\tilde{m}=3}^{n}\ \sum_{|m|=\tilde{m}}(d_{0}\|A\|)^{\tilde{m}}\
\frac{c_{6}\,c_{1}\,\nu\,D_{2}^{\tilde{m}}\,m!}{(1+{\tilde{m}})^{d+1}\,m!}\
\sqrt{\frac{(J+(n-1)+2(p-|l|-k))!}{J!}}\\[5pt] \nonumber
&&\qquad\qquad\qquad\qquad\qquad
\times\quad D_{1}\,D_{2}^{|\alpha |+|l|+4(n-1-\tilde{m})}\
\frac{(\alpha+l)!}{(1+|\alpha|)^{d+1}}\ \frac{|t|^{p}}{p!}\
\frac{{k}^{k}}{\dist(t)^{k}}\\[5pt] \nonumber
&\leq&\sum_{\tilde{m}=3}^{n}\ \sum_{|m|=\tilde{m}}D_{1}\,c_{6}\,c_{1}\nu\,
\frac{(d_{0}\|A\|)^{\tilde{m}}}{D_{2}^{3\tilde{m}}}\,
D_{2}^{|\alpha |+|l|+4(n-1)}\,
\frac{(\alpha+l)!}{(1+|\alpha|)^{d+1}}\,\frac{|t|^{p}}{p!}\,
\frac{{k}^{k}}{\dist(t)^{k}}\\[5pt] \nonumber
&&\qquad\qquad\qquad\qquad\qquad\qquad\qquad\qquad
\times\quad\sqrt{\frac{(J+n-1+2(p-|l|-k))!}{J!}}\,.
\eea
We also verify the constraints on the parameters and components
of the vectors:
\bea\nonumber
p&\leq&n-1- \tilde{m}\quad\leq\quad n-1,\\[3pt] \nonumber
|l|+k&\leq&p+(n-1-\tilde{m})/2\quad\leq\quad p+(n-1)/2,\\[3pt] \nonumber
|j|&\leq&J+(n-\tilde{m}-1)+2(p-|l|-k)+\tilde{m}\quad\le\quad J+(n-1)+2(p-|l|-k).
\eea
Hence, we see that each contribution from $\Delta_{2}(w,t)$ satisfies
the required bound, provided the following two conditions are fulfilled
\bea\nonumber
D_{2}^{3}&\geq&d_{0}\,\|A\|,\\[4pt]\nonumber
D_{2}^{9}&\geq&(d_{0}\|A\|)^{3}\,c_{6}\,c_{1}\,\nu .
\eea
There are
$$
\sum_{\tilde{m}=3}^{n}\ \sum_{|m|=\tilde{m}}\ 1\ \leq\ \sum_{|m|\leq n}\ 1
\ =\ \pmatrix{n+d\cr d}\ \leq\ \sigma_{0}\,\E^{\sigma n}
$$
such contributions, where $\sigma>0$ can be chosen arbitrarily small
(see \cite{hagjoy3}).

\vspace{.5cm}
\noindent {\bf The Term $\Delta_{3}$}\\

For this term we make the laplacian explicit and write
$$
\Delta_{w}d_{n-4,p,l,k}(w,t)\ =\ \sum_{i=1}^{d}\
(D^{2}_{w_{i}}d_{n-4,p,l,k})(w,t).
$$
We introduce\quad $\ds l_{i,2}\ =\ l+(0,0,\dots 0,2,0,\dots,0)$,
where the $2$ sits in the $i$th column.\\
We then estimate
\bea\nonumber
&&\left\|\,D_{w}^{\alpha}\ \frac{1}{2}\ \left( rP_{\perp}(w,t)
\Delta_{w}{d}_{n-4,p,l,k}(w,t)\,
\right)\,\right\|\\[5pt] \nonumber
&\leq&\frac{1}{2}\,\sum_{i=1}^{d}\,c_{1}\,\nu\,D_{1}\,
D_{2}^{|\alpha|+|l|+2+4(n-5)}\
\frac{(\alpha+l_{i,2})!}{(1+|\alpha|)^{d+1}}\
\frac{|t|^{p}}{p!}\ \frac{{k}^{k}}{\dist(t)^{k}}
\\[5pt] \nonumber
&&\quad\quad\quad\quad\quad\quad\quad\quad\quad\quad\quad\quad\quad\quad
\quad\times\quad\sqrt{\frac{(J+(n-5)+2(p-|l|-k))!}{J!}}\\[5pt] \nonumber
&=&\sum_{i=1}^{d}\ \frac{c_{1}\,\nu\,D_{1}}{2\,D_{2}^{16}}\
D_{2}^{|\alpha|+|l_{i,2}|+4(n-1)}\ \frac{(\alpha+l_{i,2})!}{(1+|\alpha|)^{d+1}}\
\frac{|t|^{p}}{p!}\ \frac{{k}^{k}}{\dist(t)^{k}}\\[5pt]\nonumber
&&\quad\quad\quad\quad\quad\quad\quad\quad\quad\quad\quad\quad\quad\quad
\quad\times\quad\sqrt{\frac{(J+(n-1)+2(p-|l_{i,2}|-k))!}{J!}}.
\eea
Again, the constraints are satisfied since
\bea\nonumber
p&\leq&n-5\quad <\quad n-1\\[3pt] \nonumber
|l_{i,2}|+k&\leq&p+(n-5)/2+2\quad=\quad p+(n-1)/2\\[3pt] \nonumber
|j|&\leq&J+(n-5)+2(p-|l|-k)\quad=\quad J+(n-1)+2(p-|l_{i,2}|-k)
\eea
and each of the $d$ contributions stemming from $\Delta_{3}(w,t)$
satisfies the required estimate, provided
$$
D_{2}^{16}\ \geq\ c_{1}\nu/2.
$$

We estimate each of the remaining terms $\Delta_{i}(w,t)$,\ $i=4,\dots, 8$,
in the same fashion, using the same tools. Since this is straightforward, we
only outline the arguments.

\vspace{.5cm}
\noindent {\bf The Term $\Delta_{4}$}\\

We expand the dot product
$$
(\nabla_{w}\Phi)\cdot(\nabla_{w}c_{n-4,p,l,k})\ =\
\sum_{i=1}^{d}\ (D_{w_{i}}\Phi)\,(D_{w_{i}}c_{n-4,p,l,k})
$$
and use the definition
$$
l_{i,1}=l+(0,0,\dots 0,1,0,\dots,0),
$$
where the $1$ sits at the $i$th column. Recall that
the estimates on the $c_{m,p,l,k}$'s differ from those on the
$d_{m,p,l,k}$'s by a shift of $1$ in the $m$ dependence.
We have
\bea\nonumber
&&\left\|\,D_{w}^{\alpha}\left( rP_{\perp}(w,t)\,
\nabla_{w}\Phi(w,t) \cdot \nabla_{w}{c}_{n-4,p,l,k}(w,t)
\right)\,\right\|\\[5pt] \nonumber
&\leq&\sum_{i=1}^{d}\ \frac{c_{1}\,c_{4}\,\nu^{2}\,D_{1}}{D_{2}^{12}}\
D_{2}^{|\alpha|+|l_{i,1}|+4(n-1)}\
\frac{(\alpha+l_{i,1})!}{(1+|\alpha|)^{d+1}}\
\frac{|t|^{p}}{p!}\ \frac{{k}^{k}}{\dist(t)^{k}}\\[5pt] \nonumber
&&\qquad\qquad\qquad\qquad\qquad\qquad\qquad\qquad\times\quad
\sqrt{\frac{(J+(n-1)+2(p-|l_{i, 1}|-k))!}{J!}},
\eea
with all constraints on $|j|,p,|l_{i,1}|,k$ satisfied.
Thus each of the $d$ contributions stemming from $\Delta_{4}(w,t)$
satisfies the required estimate, provided
$$
D_{2}^{12}\ \geq\ c_{1}\,c_{4}\,\nu^{2}.
$$

\vspace{.5cm}
\noindent {\bf The Term $\Delta_{5}$}\\

This term is similar to the previous one. We obtain
\bea\nonumber
&&\left\|\,D_{w}^{\alpha}\left(\,\frac 12\, rP_{\perp}(w,t)\,
(\Delta_{w}\Phi)(w,t)\,c_{n-4,p,l,k}(w,t)
\right)\,\right\|\\[5pt] \nonumber
&\leq&\frac{c_{1}\,c_{5}\,\nu^{2}\,D_{1}}{2D_{2}^{12}}\
D_{2}^{|\alpha |+|l|+4(n-1)}\ \frac{(\alpha+l)!}{(1+|\alpha|)^{d+1}}\
\frac{|t|^{p}}{p!}\ \frac{{k}^{k}}{\dist(t)^{k}}\\[5pt]\nonumber
&&\qquad\qquad\qquad\qquad\qquad\qquad\qquad\qquad\times\quad
\sqrt{\frac{(J+(n-1)+2(p-|l|-k))!}{J!}},
\eea
with all constraints on $|j|,p,|l|,k$ satisfied.
Thus the contribution stemming from $\Delta_{5}(w,t)$
satisfies the required estimate, provided
$$
D_{2}^{12}\ \geq\ c_{1}\,c_{5}\,\nu^{2}/2.
$$

\vspace{.5cm}
\noindent {\bf The Term $\Delta_{6}$}\\

At this point the matrices $\bra\ffi,\,D_{y_{i}}\ffi\ket$
play a role that we control by the momentum space analog of Lemma
\ref{ylem}.
Expanding the dot product and introducing the matrices
$\bra\ffi,\,D_{y_{i}}\ffi\ket$,\ $i=1,\dots,d$ we have the following estimate
for this term:
\bea\nonumber
&&\left\|\,D_{w}^{\alpha}\ \sum_{i=1}^{d}\
(rP_{\perp})(w,t)\ \bra\ffi,\,D_{y_{i}}\ffi\ket\
D_{w_{i}}d_{n-3,p,l,k}(w,t)\,\right\| \\[5pt] \nonumber
&\leq&\sum_{i=1}^{d}\ \frac{d_{0}\,c_{1}\,\nu\,\|B\|\,D_{1}}{D_{2}^{12}}\
D_{2}^{|\alpha|+|l_{i,1}|+4(n-1)}\ \frac{(\alpha+l_{i,1})!}{(1+|\alpha|)^{d+1}}\
\frac{|t|^{p}}{p!}\ \frac{{k}^{k}}{\dist(t)^{k}}\\[5pt] \nonumber
&&\qquad\qquad\qquad\qquad\qquad\qquad\qquad\qquad\times\quad
\sqrt{\frac{(J+(n-1)+2(p-|l_{i,1}|-k))!}{J!}}
\eea
with all constraints on $|j|,p,|l_{i,1}|,k$ satisfied.
Thus, each of the $d$ contributions stemming from $\Delta_{6}(w,t)$
satisfies the required estimate, provided
$$
D_{2}^{12}\ \geq\ d_{0}\,c_{1}\,\nu \|B\|.
$$

\vspace{.5cm}
\noindent {\bf The Term $\Delta_{7}$}\\

Similarly,
\bea\nonumber
&&\left\|\,D_{w}^{\alpha}\ \sum_{i=1}^{d}\
(rP_{\perp})(w,t)\ \bra\ffi,\,D_{y_{i}}\ffi\ket\
(D_{w_{i}}\Phi)(w,t)\ c_{n-3,p,l,k}(w,t)\,\right\|\\[5pt] \nonumber
&\leq&\sum_{i=1}^{d}\
\frac{d_{0}\,c_{1}\,c_{4}\,\nu^{2}\,\|B\|\,D_{1}}{D_{2}^{8}}\
D_{2}^{|\alpha|+|l|+4(n-1)}\ \frac{(\alpha+l)!}{(1+|\alpha|)^{d+1}}\
\frac{|t|^{p}}{p!}\ \frac{{k}^{k}}{\dist(t)^{k}}\\[5pt] \nonumber
&&\qquad\qquad\qquad\qquad\qquad\qquad\qquad\qquad\times\quad
\sqrt{\frac{(J+(n-1)+2(p-|l|-k))!}{J!}},
\eea
with all constraints on $|j|,p,|l|,k$ satisfied.
Thus, each of the $d$ contributions stemming from $\Delta_{7}(w,t)$
satisfies the required estimate, provided
$$
D_{2}^{12}\ \geq\ d_{0}\,c_{1}\,c_{4}\,\nu^{2}\,\|B\|.
$$

\vspace{.5cm}
\noindent {\bf The Term $\Delta_{8}$}\\

Finally,
\bea\nonumber
&&\left\|\,D_{w}^{\alpha}\,\left( rP_{\perp}(w,t)\,
\dot{\Phi}(w,t)\,c_{n-2,p,l,k}(w,t)\right)\,\right\|\\[5pt] \nonumber
&&\!\!\!\!\!\!\!\!\!\!\!\!\!\!
\leq\ \frac{c_{1}\,c_{3}\,\nu^{2}\,D_{1}}{D_{2}^{4}}\
D_{2}^{|\alpha|+|l|+4(n-1)}\ \frac{(\alpha+l)!}{(1+|\alpha|)^{d+1}}\
\frac{|t|^{p}}{p!}\ \frac{{k}^{k}}{\dist(t)^{k}}\
\sqrt{\frac{(J+(n-1)+2(p-|l|-k))!}{J!}},
\eea
with all constraints on $|j|,p,|l|,k$ satisfied.
Thus the contribution stemming from $\Delta_{8}(w,t)$
satisfies the required estimate, provided
$$
D_{2}^{4}\ \geq\ c_{1}\,c_{3}\,\nu^{2}.
$$

We now perform a similar analysis for the quantities
$\Gamma_{i}(w,t)$ that appear in the expression for $\dot{c}_{n,p,l,k}(w,t)$.
We integrate these terms with respect to $t$ and apply Lemma \ref{integrate}.
According to the lemma, integration of a term with a given value of $p$ gives
rise to a term with $p'=p+1$ in the estimates.
We also note that the estimates
we want to prove for the $c$'s differ from those for
the $d$'s by the replacement of $n-1$ by $n$.

\vspace{.5cm}
\noindent {\bf The Term $\int_{0}^{t}\,\Gamma_{1}$}\\

We use the same techniques above to obtain
\bea\nonumber
&&\left\|\,D_{w}^{\alpha}\ \int_{0}^{t}\
\frac 12\ (\Delta_{w} {c}_{n-2, p, l, k})(w,s)\, ds\,\right\|\\[5pt] \nonumber
&\leq&\sum_{i=1}^{d}\ \frac{D_{1}}{2D_{2}^{8}}\
D_{2}^{|\alpha |+|l_{i, 2}|+4n}\
\frac{(\alpha+l_{i,2})!}{(1+|\alpha|)^{d+1}}\ \frac{t^{p'}}{p'!}\
\frac{{k}^{k}}{\dist(t)^{k}}\
\sqrt{\frac{(J+n+2(p'-|l_{i, 2}|-k))!}{J!}}.
\eea
We check that the constraints are satisfied:
\bea\nonumber
p'&\leq&n-1\quad<\quad n,\\[3pt] \nonumber
|l_{i, 2}|+k&\leq&p+(n-2)/2+2\quad =\quad p'+n/2\\[3pt] \nonumber
|j|&\leq&J+(n-2)+2(p-|l|-k)\quad =\quad J+n+2(p'-|l_{i,2}|-k).
\eea
Thus, each of the $d$ contributions stemming from $\Gamma_{1}(w,t)$
satisfies the required estimate provided
$$
D_{2}^{8}\ \geq\ 1/2.
$$

\vspace{.5cm}
\noindent {\bf The Term $\int_{0}^{t}\,\Gamma_{2}$}\\

Similarly, with $p'=p+1$,
\bea\nonumber
&&\left\|\,D_{w}^{\alpha}\ \int_{0}^{t}\ \sum_{i=1}^{d}\
\bra\Phi,\,D_{w_{i}}\Phi\ket(w,s)\ D_{w_{i}}{c}_{n-2,p,l,k}(w,s)\,ds\,\right\|
\\[5pt] \nonumber
&\leq&\sum_{i=1}^{d}\ \frac{c'_{4}\,\nu\,D_{1}}{D_{2}^{8}}\
D_{2}^{|\alpha|+|l_{i,1}|+4n}\ \frac{(\alpha+l_{i,1})!}{(1+|\alpha|)^{d+1}}\
\frac{|t|^{p'}}{p'!}\ \frac{{k}^{k}}{\dist(t)^{k}}\
\sqrt{\frac{(J+n+2(p'-|l_{i, 1}|-k))!}{J!}},
\eea
with all constraints on $|j|,p',|l_{i,1}|,k$ satisfied.
Thus each of the $d$ contributions stemming from $\Gamma_{2}(w,t)$
satisfies the required estimate, provided
$$
D_{2}^{8}\ \geq\ c'_{4}\,\nu.
$$

\vspace{.5cm}
\noindent {\bf The Term $\int_{0}^{t}\,\Gamma_{3}$}\\

Again, with $p'=p+1$,
\bea\nonumber
&&\left\|\,D_{w}^{\alpha}\ \int_{0}^{t}\ \frac{1}{2}
\bra\Phi,\,(\Delta_{w}\Phi)\ket(w,s)\
c_{n-2,p,l,k}(w,s)\,ds\, \right\| \\[5pt] \nonumber
&\leq&\frac{c'_{5}\,\nu\,D_{1}}{2\,D_{2}^{8}}\
D_{2}^{|\alpha|+|l|+4n}\ \frac{(\alpha+l)!}{(1+|\alpha|)^{d+1}}\
\frac{|t|^{p'}}{p'!}\ \frac{{k}^{k}}{\dist(t)^{k}}\
\sqrt{\frac{(J+n+2(p'-|l|-k))!}{J!}},
\eea
with all constraints on $|j|,p',|l|,k$ satisfied.
Thus, the contribution stemming from $\Gamma_{3}(w,t)$
satisfies the required estimate, provided
$$
D_{2}^{8}\ \geq\ c'_{5}\,\nu/2.
$$

\vspace{.5cm}
\noindent {\bf The Term $\int_{0}^{t}\,\Gamma_{4}$}\\

Recall that the matrices $\bra\ffi,\,D_{y_{i}}\ffi\ket$
are controlled by an analog of Lemma \ref{ylem}.
\bea\nonumber
&&\left\|\,D_{w}^{\alpha}\ \int_{0}^{t}\ \sum_{i=1}^{d}\
\bra\ffi,\,D_{y_{i}}\ffi\ket\ D_{w_{i}}{c}_{n-1,p,l,k}(w,s)\,ds\,
\right\|\\[5pt] \nonumber
&&\!\!\!\!\!\!\!\!\!\!\!\!
\leq\ \sum_{i=1}^{d}\ \frac{d_{0}\,\|B\|\,D_{1}}{D_{2}^{4}}\
D_{2}^{|\alpha|+|l_{i,1}|+4n}\ \frac{(\alpha+l_{i,1})!}{(1+|\alpha|)^{d+1}}\
\frac{|t|^{p'}}{p'!}\ \frac{{k}^{k}}{\dist(t)^{k}}\
\sqrt{\frac{(J+n+2(p'-|l_{i,1}|-k))!}{J!}},
\eea
with all constraints on $|j|,p',|l_{i,1}|,k$ satisfied.
Thus each of the $d$ contributions stemming from $\Gamma_{4}(w,t)$
satisfies the required estimate, provided
$$
D_{2}^{4}\ \geq\ d_{0}\,\|B\|.
$$

\vspace{.5cm}
\noindent {\bf The Term $\int_{0}^{t}\,\Gamma_{5}$}\\

For this term we obtain
\bea\nonumber
&&\left\|\,D_{w}^{\alpha}\ \int_{0}^{t}\ \sum_{i=1}^{d}\
\bra\Phi,\,(D_{w_{i}}\Phi)(w,s)\ket\,\bra\ffi,\,D_{y_{i}}\ffi\ket\
c_{n-1,p,l,k}(w,s)\,ds\,\right\|\\[5pt] \nonumber
&&\!\!\!\!\!\!\!\!\!\!\!\!\!
\leq\ \sum_{i=1}^{d}\ \frac{c'_{4}\,\nu\,d_{0}\,\|B\|\,D_{1}}{D_{2}^{4}}\
D_{2}^{|\alpha|+|l|+4n}\ \frac{(\alpha+l)!}{(1+|\alpha|)^{d+1}}\
\frac{|t|^{p'}}{p'!}\ \frac{{k}^{k}}{\dist(t)^{k}}\
\sqrt{\frac{(J+n+2(p'-|l|-k))!}{J!}},
\eea
with all constraints on $|j|,p',|l|,k$ satisfied.
Thus each of the $d$ contributions stemming from $\Gamma_{5}(w,t)$
satisfies the required estimate, provided
$$
D_{2}^{4}\ \geq\ c'_{4}\,\nu\,d_{0}\,\|B\|.
$$

\vspace{.5cm}
\noindent {\bf The Term $\int_{0}^{t}\,\Gamma_{6}$}\\

In this term, we encounter the sum over all previous $c$'s.
As in the similar contribution from $\Delta_{2}$, we obtain
\bea\nonumber
&&\left\|\,D_{w}^{\alpha}\ \sum_{\tilde{m}=3}^{n+2}\ \sum_{|m|=\tilde{m}}\
\int_{0}^{t}\,\frac{D^{m}E(a(t))}{m!}\ \bra\ffi,\,y^{m}\ffi\ket\
c_{n+2-\tilde{m},p,l,k}(w,s)\, \right\| \\[5pt] \nonumber
&\leq&\sum_{\tilde{m}=3}^{n}\ \sum_{|m|=\tilde{m}}\ D_{1}\ c_{6}\ D_{2}^{8}\
\frac{(d_{0}\,\|A\|)^{\tilde{m}}}{D_{2}^{3\tilde{m}}}\
D_{2}^{|\alpha|+|l|+4n}\
\frac{(\alpha+l)!}{(1+|\alpha|)^{d+1}}\ \frac{|t|^{p'}}{p'!}\
\frac{{k}^{k}}{\dist(t)^{k}}\\[5pt] \nonumber
&&\qquad\qquad\qquad\qquad\qquad\qquad\qquad\times\quad
\sqrt{\frac{(J+n+2(p'-|l|-k))!}{J!}}.
\eea
We check that the constraints on the parameters and components
of the vectors are satisfied
\bea\nonumber
p'&\leq&n-\tilde{m}+3\quad\leq\quad n\\[3pt] \nonumber
|l|+k&\leq&p+(n-\tilde{m}+2)/2\quad\leq\quad p'+n/2\\[3pt] \nonumber
|j|&\leq&J+n+2+2(p-|l|-k)\quad =\quad J+n+2(p'-|l|-k).
\eea
Hence we see that each contribution from $\Delta_{2}(w,t)$ satisfies
the required bound, provided the following two conditions are fulfilled
\bea\nonumber
D_{2}^{3}&\geq&d_{0}\ \|A\| \\[3pt] \nonumber
D_{2}&\geq&(d_{0}\,\|A\|)^{3}\,c_{6}.
\eea
There are
$\ds
\sum_{\tilde{m}=3}^{n}\ \sum_{|m|=\tilde{m}}\ 1\ \leq\ \sigma_{0}\,\E^{\sigma n}
$
such contributions, where $\sigma >0$ can be chosen arbitrarily small.

\vspace{.5cm}
\noindent {\bf The Term $\int_{0}^{t}\,\Gamma_{7}$}\\

This terms depends on the $d$'s. Recall the estimates are
a little different for them.
\bea\nonumber
&&\left\|\,D_{w}^{\alpha}\ \int_{0}^{t}\ \sum_{i=1}^{d}\ \frac 12\
\bra\Phi(w,s),\,(D^{2}_{w_{i}}{d}_{n-2,p,l,k})(w,s)\ket\,ds\,
\right\| \\[5pt] \nonumber
&\leq&\sum_{i=1}^{d}\ \frac{c_{2}\,\nu\,D_{1}}{2\,D_{2}^{12}}\
D_{2}^{|\alpha|+|l_{i,2}|+4n}\ \frac{(\alpha+l_{i,2})!}{(1+|\alpha|)^{d+1}}\
\frac{|t|^{p'}}{p'!}\ \frac{{k}^{k}}{\dist(t)^{k}}\
\sqrt{\frac{(J+n+2(p'-|l_{i,2}|-k))!}{J!}},
\eea
with all constraints on $|j|,p',|l_{i,2}|,k$ satisfied.
Thus each of the $d$ contributions stemming from $\Gamma_{7}(w,t)$
satisfies the required estimate, provided
$$
D_{2}^{12}\ \geq\ c_{2}\,\nu/2.
$$

\vspace{.5cm}
\noindent {\bf The Term $\int_{0}^{t}\,\Gamma_{8}$}\\

Similarly,
\bea\nonumber
&&\left\|\,D_{w}^{\alpha}\ \int_{0}^{t}\ \sum_{i=1}^{d}\
\bra\ffi,\,D_{y_{i}}\ffi\ket\
\bra\Phi(w,s),\,(D_{w_{i}}{d}_{n-1,p,l,k})(w,s)\ket\,ds\,
\right\| \\[5pt] \nonumber
&&\!\!\!\!\!\!\!\!\!\!\!\!
\!\!\!\!\!\!\!\!\!\!\!\!
\leq\ \sum_{i=1}^{d}\ \frac{c_{2}\,\nu\,d_{0}\,\|B\|\,D_{1}}{D_{2}^{8}}\
D_{2}^{|\alpha|+|l_{i,1}|+4n}\ \frac{(\alpha+l_{i,1})!}{(1+|\alpha|)^{d+1}}\
\frac{|t|^{p'}}{p'!}\ \frac{{k}^{k}}{\dist(t)^{k}}\
\sqrt{\frac{(J+n+2(p'-|l_{i, 1}|-k))!}{J!}},
\eea
with all constraints on $|j|,p',|l_{i,1}|,k$ satisfied.
Thus, each of the $d$ contributions stemming from $\Gamma_{8}(w,t)$
satisfies the required estimate provided
$$
D_{2}^{8}\ \geq\ c_{2}\,\nu\,d_{0}\,\|B\|.
$$

\vspace{.5cm}
\noindent {\bf The Term $\int_{0}^{t}\,\Gamma_{9}$}\\

Finally,
\bea\nonumber
&&\left\|\,D_{w}^{\alpha}\ \int_{0}^{t}\
\bra\dot{\Phi}(w,s),\,d_{n,p,l,k}(w,s)\ket\,ds\,\right\|\\[5pt] \nonumber
&\leq&\frac{c_{3}\,\nu\,D_{1}}{D_{2}^{4}}\ D_{2}^{|\alpha |+|l|+4n}\
\frac{(\alpha+l)!}{(1+|\alpha|)^{d+1}}\ \frac{|t|^{p'}}{p'!}\
\frac{{k}^{k}}{\dist(t)^{k}}\ \sqrt{\frac{(J+n+2(p'-|l|-k))!}{J!}},
\eea
with all constraints on $|j|,p',|l|,k$ satisfied.
Thus, the contribution stemming from $\Gamma_{9}(w,t)$
satisfies the required estimate, provided
$$
D_{2}^{4}\ \geq\ c_{3}\,\nu .
$$

By choosing $D_2$ large enough, all conditions are satisfied.
This completes the induction for part iii) of Proposition \ref{mainprop}.

The integration required to construct the $c$'s shows that we obtain non-zero
results for $c_{n,p,l,k,\beta}$ for $n>0$ only when $p\geq 1$.
This proves part i) of Proposition \ref{mainprop}.

\vskip .3cm
We now turn to the proof of part ii) of Proposition \ref{mainprop}.

\vskip .5cm
\subsection{Counting the Number of Terms that Occur in Our Expansion}

In our Born--Oppenheimer expansion, the $n^{\mbox{\scriptsize th}}$ order
term has the form\\
$\phi_n(w,y,t)\,=\,g_n(w,y,t)\Phi(w,t)\,+\,\phi_n^{\perp}(w,y,t)$.
The way we compute $g_n(w,y,t)$ and $\phi_n^{\perp}(w,y,t)$, they
decompose naturally as sums over the parameter $\beta$. We define $u_n$
to be the number of such terms in $g_n(w,y,t)$ and $v_n$ to be the number
of terms in $\phi_n^{\perp}(w,y,t)$.

An examination of our construction shows that $u_n$ and $v_n$ satisfy
the recursive estimates
\bea\label{uinequality}
u_{n+1}&\le&\sum_{j=0}^3\,a_j\,u_{n-j}\,+\,
\sum_{j=0}^3\,b_j\,v_{n-j}\,+\,\sum_{j=0}^n\,c_1\,\gamma_1^j\,u_{n-j}
\,+\,\sum_{j=0}^n\,c_2\,\gamma_2^j\,v_{n-j}\,+\,v_{n+1}\\
v_{n+1}&\le&\sum_{j=0}^3\,d_j\,u_{n-j}\,+\,
\sum_{j=0}^3\,e_j\,v_{n-j}\,+\,\sum_{j=0}^n\,c_3\,\gamma_3^j\,u_{n-j}
\,+\,\sum_{j=0}^n\,c_4\,\gamma_4^j\,v_{n-j},\label{vinequality}
\eea
where $a_i$, $b_i$, $c_i$, $d_i$, $e_i$ and $\gamma_i$ are fixed numbers.
The exponentials $\gamma_i^j$ arise from an estimate
(proven in the proof of Lemma 5.2 of \cite{hagjoy5})
for the number of Taylor series terms of any given
order in the expansion of $E(a(t)+\epsilon y)$.

We substitute (\ref{vinequality}) for the last term in (\ref{uinequality})
and add the result to (\ref{vinequality}). By some simple estimates this
leads to a recursive estimate for the single quantity $z_n=u_n+v_n$ of the
form
$$z_{n+1}\ \le\ \sum_{j=0}^3\ \widetilde{a}_j\,z_{n-j}\ +\
\sum_{j=0}^n\ \widetilde{c}\ \widetilde{\gamma}^j\,z_{n-j}.$$

An easy induction on $n$ shows that this implies that $z_n$ grows at
most like $e^{kn}$ for a sufficiently large value of $k$.
The quantity $z_n$ is the number of terms in $\phi_n(w,y,t)$, so this
proves the assertion.\ep

\vskip .3cm
Proposition \ref{mainprop} now follows easily. \ep

\vskip .5cm
\section{Exponential Error Bounds}
\setcounter{equation}{0}

In this section, we prove Theorem \ref{main}.

\subsection{The Explicit Error Term}

We use the following abstract lemma, whose proof
is an easy application of Duhamel's formula (see e.g. \cite{raise}).

\vskip .25cm
\noindent
\begin{lem}\label{magic}
Suppose $H(\hbar)$ is a family of self-adjoint
operators for $\hbar>0$. Suppose $\psi(t,\,\hbar)$ belongs to the domain
of $H(\hbar)$, is continuously differentiable in $t$, and approximately
solves the Schr\"odinger equation in the sense that
$$
i\,\hbar\,\frac{\partial\psi}{\partial t}(t,\,\hbar)\ =\
H(\hbar)\,\psi(t,\,\hbar)\ +\ \xi(t,\,\hbar),
$$
where $\xi(t,\,\hbar)$ satisfies
$$
\|\,\xi(t,\,\hbar)\,\|\ \le \,\mu(t,\,\hbar).
$$
Then, for $t>0$,
$$
\|\,\E^{-itH(\hbar)/\hbar}\,\psi (0,\,\hbar)\ -\ \psi(t,\,\hbar)\,\|\
\le\ \hbar^{-1}\ \int_0^{t}\,\mu(s,\,\hbar)\,ds.
$$
The analogous statement holds for $t<0$.
\end{lem}

\vskip .4cm
We substitute our approximate solution (\ref{approx})\\
$\ds F\ \E^{iS/\eps^2}\ \E^{i\eta\cdot y/\eps}\
\left(\,\sum_{n=0}^N\,\epsilon^n\,\phi_n\,+\,\eps^{N+1}\,\phi_{N+1}^\perp\,+\,
\eps^{N+2}\,\phi_{N+2}^\perp\,\right)$
into the Schr\"odinger equation and compute the residual term $\xi_N$.

It is more convenient to write this term in the multiple scales notation.
We also use the notation
$\ds \eps^m\,\frac{E^{(m)}(a(t))}{m!}y^m$ to denote the Taylor series term
$\ds \sum_{|j|=m}\,\eps^{|j|}\,\frac{(D^jE)(a(t))}{j!}\,y^j$.

In this notation, the residual $\xi_N(w,y,t)$ is given, up to a phase factor,
by two sums of terms. The first one contains all terms that do not involve
derivatives of the cut-off. The second contains all terms that do
involve derivatives of the cut-off.\\
The first sum is $F(w)$ times the following:
\bea\label{errterm}
&&\frac{\eps^{N+3}}{2}\,\left(\Delta_wg_{N-1}\right)\,\Phi
\\[7pt] 
&+&\frac{\eps^{N+4}}{2}\,\left(\Delta_wg_{N}\right)\,\Phi
\\[7pt] 
&+&\eps^{N+3}\,\left(\nabla_wg_{N-1}\right)\cdot\left(\nabla_w\Phi\right)
\\[7pt] 
&+&\eps^{N+4}\,\left(\nabla_wg_{N}\right)\cdot\left(\nabla_w\Phi\right)
\\[7pt] 
&+&\frac{\eps^{N+3}}{2}\,g_{N-1}\,\left(\Delta_w\Phi\right)
\\[7pt] 
&+&\frac{\eps^{N+4}}{2}\,g_{N}\,\left(\Delta_w\Phi\right)
\\[7pt] 
&+&\frac{\eps^{N+3}}{2}\,\left(\Delta_w\phi_{N-1}^\perp\right)
\\[7pt] 
&+&\frac{\eps^{N+4}}{2}\,\left(\Delta_w\phi_{N}^\perp\right)
\\[7pt] 
&+&\eps^{N+3}\,\left(\nabla_w\cdot\nabla_yg_{N}\right)\Phi
\\[7pt] 
&+&\eps^{N+3}\,\left(\nabla_yg_{N}\right)\cdot\left(\nabla_w\Phi\right)
\\[7pt] 
&+&\eps^{N+3}\,\left(\nabla_w\cdot\nabla_y\phi_{N}^\perp\right)
\\[7pt] 
\label{dot1}
&+&i\,\eps^{N+3}\,\dot{\phi}_{N+1}^\perp
\\[7pt] 
\label{dot2}
&+&i\,\eps^{N+4}\,\dot{\phi}_{N+2}^\perp
\\[7pt] 
&+&\frac{\eps^{N+5}}{2}\,\left(\Delta_w\phi_{N+1}^\perp\right)
\\[7pt] 
&+&\frac{\eps^{N+6}}{2}\,\left(\Delta_w\phi_{N+2}^\perp\right)
\\[7pt] 
&+&\eps^{N+4}\,\left(\nabla_w\cdot\nabla_y\phi_{N+1}^\perp\right)
\\[7pt] 
&+&\eps^{N+5}\,\left(\nabla_w\cdot\nabla_y\phi_{N+2}^\perp\right)
\\[7pt] 
&+&\frac{\eps^{N+3}}{2}\,\left(\Delta_y\phi_{N+1}^\perp\right)
\\[7pt] 
&+&\frac{\eps^{N+4}}{2}\,\left(\Delta_y\phi_{N+2}^\perp\right)
\\[7pt] 
&-&\frac{\eps^{N+3}}{2}\,E^{(2)}(a(t))\,y^2\,\phi_{N+1}^\perp
\\[7pt] 
&-&\frac{\eps^{N+4}}{2}\,E^{(2)}(a(t))\,y^2\,\phi_{N+2}^\perp
\\[7pt] 
\label{sumerr1}
&-&\sum_{n=0}^N\,\eps^{N-n}\,\left(E(a(t)+\eps y)\,-\,
\sum_{m\le 2+n}\,\eps^m\,\frac{E^{(m)}(a(t))}{m!}\,y^m\,
\right)\,g_{N-n}\,\Phi
\\[7pt] 
\label{sumerr2}
&-&\sum_{n=0}^N\,\eps^{N-n}\,\left(E(a(t)+\eps y)\,-\,
\sum_{m\le 2+n}\,\eps^m\,\frac{E^{(m)}(a(t))}{m!}\,y^m\,
\right)\,\phi_{N-n}^\perp
\\[7pt] 
&-&\eps^{N+1}\,\left(E(a(t)+\eps y)\,-\,
\sum_{m\le 2}\,\eps^m\,\frac{E^{(m)}(a(t))}{m!}\,y^m\,
\right)\,\phi_{N+1}^\perp
\\[7pt]\label{moose}
&-&\eps^{N+2}\,\left(E(a(t)+\eps y)\,-\,
\sum_{m\le 2}\,\eps^m\,\frac{E^{(m)}(a(t))}{m!}\,y^m\,
\right)\,\phi_{N+2}^\perp .
\eea

The second sum arises from terms in which the cut-off $F(w)$ is differentiated.
It is
\bea\label{cutoff1}
&&\sum_{n=0}^N\ \frac{\eps^{n+4}}2\ (\Delta_wF)\,g_n\,\Phi
\\[7pt] 
&+&\sum_{n=0}^{N+2}\ \frac{\eps^{n+4}}2\ (\Delta_wF)\,\phi_n^\perp
\\[7pt] 
&+&\sum_{n=0}^N\ \eps^{n+4}\ (\nabla_wF)\cdot(\nabla_wg_n)\,\Phi
\\[7pt] 
&+&\sum_{n=0}^N\ \eps^{n+4}\ g_n\,(\nabla_wF)\cdot(\nabla_w\Phi)
\\[7pt] 
&+&\sum_{n=0}^{N+2}\ \eps^{n+4}\ (\nabla_wF)\cdot(\nabla_w\phi_n^\perp)
\\[7pt] 
&+&\sum_{n=0}^N\ \eps^{n+3}\ (\nabla_wF)\cdot(\nabla_yg_n)\,\Phi
\\[7pt] \label{cutoff2}
&+&\sum_{n=0}^{N+2}\ \eps^{n+3}\ (\nabla_wF)\cdot(\nabla_y\phi_n^\perp)
\eea

\vskip .5cm
\subsection{Optimal Truncation}\label{optimal}

Each error term in the first sum (\ref{errterm})--(\ref{moose}) can be written
as a uniformly bounded function times one of the following two forms:
\bea\nonumber
{\cal A}&=&\Psi(w,t)\ \sum_{r}\ \sum_{|j|\leq\rho(r)}\ c_{r,j}(w,t)\
\ffi_{j}(y,t)\\[4pt]\nonumber
{\cal B}&=&\sum_{r'}\ \sum_{|j|\leq\rho'(r')}\ d_{r',j}(w,t)\
\ffi_{j}(y,t),
\eea
where $\Psi(w,t)\in {\cal H}_{\mbox{\scriptsize el}}$, 
$\ffi_j(y,t)=\eps^{-d/2}\ffi_j(A(t), B(t), 1, 0, 0, y)$,
$r,\,r'$ denote a collective set of indices that belong to
some finite set, and $\rho(r)$ and $\rho'(r')$ limit the number of
multi-indices $j$ allowed in the second sum.

The error term $\xi(w,y,t)\in{\cal H}_{\mbox{\scriptsize el}}$
needs to be estimated for $t\in\R$, in the following norm
\bea\nonumber
\|\xi(t)\|&=&\left\{\,\int_{\smallR^d}\
\|\xi\left(x-a(t),\,(x-a(t))/\eps,\,t\right)\,
\|^2_{{\cal H}_{\mbox{\scriptsize el}}}\,dx\,\right\}^{1/2}\\[4pt]\nonumber
&=&\left\{\,\int_{\smallR^d}\
\|\xi(w,w/\eps,t)\|^2_{{\cal H}_{\mbox{\scriptsize el}}}\,dw\,\right\}^{1/2}.
\eea

With that norm, using the Cauchy--Schwarz inequality and the
$L^2(\R^d)$ orthonormality of the $\ffi_j(y,t)$, we obtain the following
estimate for the norm of ${\cal A}$ in terms of the norm of vector
$c_r(w,t)\in l^2(\N^d,\C)$,
\be\label{norma}
\|{\cal A}\|\ \leq\ \sum_{r}\,\sup_{w\in\mbox{\scriptsize supp}\,F}\
\|\Psi(w,t)\|_{{\cal H}_{\mbox{\scriptsize el}}}
\,\sup_{w\in\mbox{\scriptsize supp}\,F}\
\|c_r(w,t)\|\,\left(\,\sum_{|j|\leq\rho(r)}\,1\,\right)^{1/2}.
\ee
By similar arguments we get the following estimate for the norm of
${\cal B}$ in terms of the norm of the vector
$d_{r'}(w,t)\in l^2(\N^d,{\cal H}_{\mbox{\scriptsize el}})$,
\be\label{normb}
\|{\cal B}\|\ \leq\ \sum_{r'}\,\sup_{w\in \mbox{\scriptsize supp}\,F}
\|d_{r'}(w,t)\|\,\left(\,\sum_{|j|\leq\rho'(r')}\,1\,\right)^{1/2}.
\ee
Note also that
\be\label{numbindic}
\sum_{|j|\leq\rho'(r')}\,1\ \leq\ \pmatrix{\rho'(r')+d\cr d},
\ee
which grows at most polynomially with $\rho'(r')$.

\begin{lem}\label{nbehave}
For $t\in [0,\,T]$, and for any $\alpha\in\N^d$ and $\gamma\in\N^d$, there
exist $C_0>0$ and $\tau_0>0$, such that
\bea\nonumber
&&\sum_{\beta\in{\cal B}_{n,1}}\ \sum_{p\leq n}\ \sum_{k+|l|\leq p+\frac n2}\
\sum_{|j|\leq J+n +2(p-|l|-k)}\left\| \Psi(w, t) D_w^\alpha D_y^\gamma
c_{n,p,l,k,\beta,j}(w,t)
\ffi_j(y,t)\right\|\\[5pt]\label{goat}
&\leq&C_0\left\{n^{1/2}\tau_0\right\}^{n}
\eea
and
\bea\nonumber
&&\sum_{\beta\in{\cal B}_{n,2}}\ \sum_{p\leq n-1}\
\sum_{k+|l|\leq p+\frac{n-1}{2}}\ \sum_{|j|\leq J+n-1+2(p-|l|-k)}
\left\|D_w^\alpha D_y^\gamma d_{n, p, l, k, \beta, j}(w,t)
\ffi_j(y,t)\right\|\\[5pt]\label{sheep}
&\leq&C_0\left\{n^{1/2}\tau_0\right\}^{n}
\eea
If the operator $D^\gamma_y$ is replaced by the operator $y^\gamma$,
the same bounds are valid.
\end{lem}

\noindent
{\bf Proof:}\quad
We begin with (\ref{goat}). We have
\bea
&&\sum_{|j|\leq J+n+2(p-|l|-k)}D_w^\alpha
c_{n, p, l, k, \beta, j}(w,t)\ D_y^\gamma \ffi_j(y,t)\nonumber\\[5pt]
&=&\sum_{|\tilde{k}|\leq J+|\gamma|+n+2(p-|l|-k)}
\left(\,\bra\ffi,\,D_y^\gamma\ffi\ket\
D_w^\alpha c_{n, p, l, k, \beta}(w,t)\right)_{\tilde{k}}\,
\ffi_{\tilde{k}}(y,t).\nonumber
\eea
We know that the vector
$\bra\ffi,\,D_y^\gamma\ffi\ket\ D_w^\alpha c_{n,p,l,k,\beta}(w,t)$
satisfies the estimate
\bea\label{replace}
&&\left\|\,\bra\ffi,\,D_y^\gamma\ffi\ket\,
D_w^\alpha c_{n,p,l,k,\beta}(w,t)\,\right\|\ \leq\
D_1\,D_2^{|\alpha|+|l|+4n}\ \frac{(\alpha+l)!}{(1+|\alpha|)^{d+1}}\
\frac{|t|^p}{p!}\ \frac{k^k}{\delta^k}\ (\|B\|\,d_0)^{|\gamma|}\nonumber\\[5pt]
&&\quad\quad\quad\quad\quad\quad\quad\quad\quad\quad\quad\quad\quad
\quad\quad\times\quad
\sqrt{\frac{(J+|\gamma|+n+2(p-|l|-k))!}{J!}}.
\eea
Here $\delta>0$ is the distance in the complex plane from $[0,\,T]$ to the
complement of $\Omega$.

Since the number of indices in ${\cal B}_{n,1}$ is bounded by
$\E^{K_0 n}$,\ \, $D_2\geq 1$,\ \, and $(\alpha+l)!\leq (|\alpha|+|l|)!$,
we can estimate the sum (\ref{goat}) by
\bea
&&\frac{D_1\,D_2^{|\alpha|}\,(\|B\|d_0)^{|\gamma|}}
{\sqrt{J!}\,(1+|\alpha|)^{d+1}}\ \E^{K_0 n}\ D_2^{11 n/2}\
\sum_{p\leq n}\ \frac{|t|^p}{p!}\
\sum_{|l|+k\leq p+n/2}\left(\frac{k}{\delta}\right)^k\nonumber\\[5pt]
&&\qquad\qquad\qquad\quad\quad\times\quad(|\alpha|+|l|)!\
\sqrt{(J+|\gamma|+n+2(p-|l|-k))!}.  \label{above}
\eea
Then, using $a!b!\leq (a+b)! $, the fact that $(a+2p)!/(p!)^2$ is
increasing in $p$, and $p\leq n$, we have
$$
\frac{(J+|\gamma|+n+2(p-|l|-k))!\ ((|\alpha|+|l|)!)^2}
{(p!)^2}\ \leq\ \frac{(J+|\gamma|+2|\alpha|+3n-2k)!}{(n!)^2},
$$
so that (\ref{above}) is bounded by
\bea\nonumber
&&\frac{D_1\,D_2^{|\alpha|}\,(\|B\|d_0)^{|\gamma|}}
{n!\,\sqrt{J!}\,(1+|\alpha|)^{d+1}}\ \E^{K_0 n}\ D_2^{11 n/2}\
\sum_{p\leq n}\ |t|^p\ \sum_{k=0}^{p+n/2}\
\left(\frac{k}{\delta}\right)^k\\[5pt]\nonumber  
&&\qquad\qquad\qquad\qquad\qquad\qquad\qquad\times\quad
\sqrt{(J+|\gamma|+2|\alpha|+3n-2k)!}\ \sum_{|l|\leq p+n/2-k}\ 1.
\eea
The last term is bounded by $\pmatrix{\grintl 3n/2\grintr +d\cr d}\,\leq\,
\sigma_0\,\e^{3\sigma n/2}$, where $\grintl x\grintr$ denotes the integer part
of $x$.

Using $k^{2k}\leq (2k)^{2k}$, $a^ab^b\leq
(a+b)^{a+b}$ and $a!\leq a^a$ we have
$$
(J+|\gamma|+2|\alpha|+3n-2k)!\ k^{2k}\ \leq\
(J+|\gamma|+2|\alpha|+3n)^{J+|\gamma|+2|\alpha|+3n}.
$$
Since we can assume without loss that $\delta<1$, this implies
\bea
&&\sum_{k=0}^{p+n/2}\,
\left(\frac{k}{\delta}\right)^k\,
\sqrt{(J+|\gamma|+2|\alpha|+3n-2k)!}\nonumber\\
&\leq&(J+|\gamma|+2|\alpha|+3n)^{\frac{J+|\gamma|+2|\alpha|+3n}{2}}\quad
\sum_{k=0}^{p+n/2}\,\delta^{-k}\nonumber \\[5pt]
&\leq&(J+|\gamma|+2|\alpha|+3n)^{\frac{J+|\gamma|+2 |\alpha| +3n}{2}}\
\frac{K_1}{\delta^{n/2 }}\ \delta^{-p},\label{45}
\eea
for some constant $K_1$ that satisfies $\delta^{-1}-1\geq K_1^{-1}\delta^{-1}$.
Together with
\be\label{46}
\sum_{p\leq n}\ (t/\delta)^p\ \leq\
\left\{\,\matrix{K_2 (t/\delta)^n&\mbox{if}&t/\delta>1\cr
                            K_2^n&\mbox{if}&t/\delta=1\cr
                              K_2&\mbox{if}&t/\delta<1,}\right.
\ee
where $K_2$ is constant, we get (in the first case above)
\bea\nonumber 
&&\sum_{\beta\in {\cal B}_{n,1}}\ \sum_{p\leq n}\
\sum_{k+|l|\leq p+\frac n2}\
\left\|\,\bra\ffi,\,D_y^\gamma\ffi\ket\,
D_w^\alpha c_{n,p,l,k,\beta}(w,t)\,\right\| \\[4pt]\nonumber
&\leq&\frac{\sigma_0K_1K_2D_1 D_2^{|\alpha|}(\|B\|d_0)^{|\gamma|}}
{(1+|\alpha|)^{d+1}}\
\frac{\E^{(K_0+3\sigma/2 )n}D_2^{11 n/2}t^n}{\sqrt{J!}\,n!\,\delta^{3n/2}}\
(J+|\gamma|+2|\alpha|+3n)^{\frac{J+|\gamma|+2|\alpha|+3n}{2}}.
\eea
We postpone the study of the dependence of our estimates on $t$ and $J$
to Section \ref{gene}. So, using the above,
$$
(J+|\gamma|+2|\alpha|+3n)^{\frac{J+|\gamma|+2|\alpha|+3n}{2}} \leq
(J+|\gamma|+2|\alpha|+3n)^{\frac{J+|\gamma|+2|\alpha|}{2}}
((J+|\gamma|+2|\alpha|+3)n)^{\frac{3n}{2}},
$$
and the existence of $0<a<b$, such that $a^nn^n\leq n!\leq b^nn^n$,
we learn the existence of positive
constants ({\it i.e.}, independent of $n$) $K_3$, $K_4$ and $K_5$, such that
$$
\sum_{\beta\in{\cal B}_{n,1}}\ \sum_{p\leq n}\ \sum_{k+|l|\leq p+\frac n2}\,
\left\|\,\bra\ffi,\,D_y^\gamma\ffi\ket\
D_w^\alpha c_{n,p,l,k,\beta}(w,t)\,\right\|\ \leq\
K_3\,K_4^{n}\ \frac{n^{3n/2}}{a^nn^n}\ \leq\ K_3\,(K_5\,n^{1/2})^n.
$$
This yields the result with $C_0=K_3$ and $\tau_0=K_5$.

The second sum is dealt with in the same manner, since the
vectors $d_{n,p,l,k,\beta}(w,t)$ satisfy the same bounds as
$c_{n,p,l,k,\beta}(w,t)$ does with $n$ replaced by $n-1$.

Finally, the replacement of $D^\gamma_y$ by the operator $y^\gamma$
means that the matrix $\bra \ffi |D^\gamma_y\ffi\ket $ must be replaced
by the matrix $\bra \ffi |y^\gamma\ffi\ket $. But the latter has the same
properties as the former; the bounds above remain true with
$\|B\|$ replaced by $\|A\|$ from (\ref{replace}) onward. This affects
the definition of $C_0$ only.\ep

\vskip .3cm
The following lemma is the key to the proof of exponential accuracy of our
approximation by means of optimal truncation.

\begin{lem}\label{optrunc}
For sufficiently small $g>0$, there exist $\Gamma(g)$ and $C(g)>0$ such that the
choice $N(\eps)=\grintl g^2/\eps^2\grintr$ implies that the norm of the error
term $\xi_{N(\eps)}(t)$ given by (\ref{errterm}) satisfies
$$\|\xi_{N(\eps)}(t)\|\ \leq\ C(g)\,\E^{-\Gamma(g)/\eps^2}.$$
\end{lem}

\noindent{\bf Proof:}\quad
The previous lemma, formulas (\ref{norma}), (\ref{normb}) and
(\ref{numbindic}) show that all terms in the first sum defining 
$\xi_N$ except (\ref{dot1}),
(\ref{dot2}), (\ref{sumerr1}), and (\ref{sumerr2}) are exponentially small,
once we prove
\be
C_0\eps^{N(\eps)}\ \left\{N(\eps)^{1/2}\ \tau_0\right\}^{N(\eps)}\ \le\
C\ \E^{-\Gamma/\eps^2}.
\ee
Because $g^2/\eps^2-1 \leq N\leq g^2/\eps^2$, if we choose
$0<g<1/\tau_0$, the left hand side of this inequality is bounded by
\bea\label{analysis}
C_0\left\{\eps N^{1/2}\tau_0\right\}^{N}\ \leq\
C_0\left\{g\tau_0\right\}^{N}\ \leq\
C_0\,\E^{-|\ln(g\tau_0)|N}\ \leq\
C_0\,\E^{|\ln(g\tau_0)|}\ \E^{-|\ln(g\tau_0)|g^2/\eps^2},
\eea
which gives
$$ C(g)\ =\ C_0\,\E^{|\ln(g\tau_0)|}\quad\mbox{and}\quad
\Gamma(g)\ =\ |\ln(g\tau_0)|\ g^2.$$
The terms (\ref{dot1}) and (\ref{dot2}) can be dealt with in a similar
fashion once we have computed
\bea
\dot\phi_{N+1}^\perp&=&\sum_{\beta\in{\cal B}_{N+1,2}}\ \sum_{p\leq N}\
\sum_{k+|l|\leq p+\frac{N}{2}}\ \,\sum_{|j|\leq J+N+2(p-|l|-k)}\
\dot d_{N+1,p,l,k,\beta,j}(w,t)\,\ffi_j(y,t)\nonumber\\[5pt]\nonumber
&&\qquad\qquad\qquad\qquad\qquad+\quad d_{N+1,p,l,k,\beta,j}(w,t)\,
\dot\ffi_j(y,t),
\eea
where the second term equals
\bea\nonumber
&&\!\!\!\!\!\!\!\!\!\!\!\!
\sum_{\beta\in{\cal B}_{N+1,2}}\ \sum_{p\leq N}\
\sum_{k+|l|\leq p+\frac{N}{2}}\ \sum_{|\tilde{k}|\leq J+N+2+2(p-|l|-k)}\
\left(\,\left(\,\frac i2\ \bra\ffi,\,\Delta_y\ffi\ket\right.\right.\\[5pt]
&&\nonumber\qquad\qquad\qquad\qquad\qquad\qquad
\left.\left.-\quad\frac {iE^{(2)}(a(t))}{2}\
\bra\ffi,\,y^2\ffi\ket\right)\,d_{N+1,p,l,k,\beta}(w,t)\right)_{\tilde k}
\,\ffi_{\tilde k}(y,t).\nonumber
\eea
Lemma \ref{upgrade} shows that $\dot d_{N+1,p,l,k,\beta}$
satisfies  bounds similar to those satisfied by $ d_{N+1,p,l,k,\beta,j}$
and the term above is taken care of by lemma
\ref{nbehave}. Similar statements are true for $\dot\phi_{N+2}^\perp$,
and the analysis above also applies to these error terms.

Next consider (\ref{sumerr1}). By the mean value theorem, there exists
$\zeta_q(y,t,\eps)=a(t)+\theta_q(y,t,\eps)\,\eps y$, where $q\in\N^d$ and
$\theta_q(y,t,\eps)\in (0,1)$, such that
$$
E(a(t)+\eps y)\,-\,
\sum_{m\le 2+n}\,\eps^m\ \frac{E^{(m)}(a(t))}{m!}\,y^m\ =\
\sum_{|q|=2+n+1}\eps^{|q|}\ \frac{D^q\,E(\zeta_q(y,t,\eps))}{q!}\ y^q.
$$
Hence, we need to estimate
\be\label{need}
\sum_{n=0}^N\ \eps^{N+3}\,\sum_{|q|=2+n+1}\,\frac{D^qE(\zeta_q(y,t,\eps))}{q!}
\ y^q\ \sum_{\beta,p,k,l,j}\,c_{N-n,p,l,k,\beta,j}(w,t)\,\ffi_j(y,t)\,\Phi,
\ee
with the following restrictions
\bea\label{restr}
|j|&\leq &J+(N-n)+2(p-k-|l|)\nonumber\\
k+|l|&\leq&p+(N-n)/2\nonumber\\
p&\leq&N-m\nonumber\\
\beta&\in&{\cal B}_{1,N-n}.
\eea

We take a fixed value of $n\in [0, N]$, and consider the vectors
$$
\frac{D^qE(\zeta_q(y,t,\eps))}{q!}\,\bra\ffi,\,y^q\ffi\ket(t)
c_{N-n,p,l,k,\beta}(w,t)
$$
we have to estimate. Due to the presence of the cut-off function $F$
(which we have omitted in the notation), we have
$$
\frac{|D^qE(\zeta_q(y,t,\eps))|}{q!}\ \leq\
\frac{c_6\,D_2^{|q|}}{(1+|q|)^{d+1}},
$$
and with our bounds on the matrix $\bra\ffi,\,y^q\ffi\ket(t)$ and on the vector
$c_{N-n,p,l,k,\beta}(w,t)$, we
can write
\bea\label{1064}
&&\left\|\,\frac{D^q\,E(\zeta_q(y,t,\eps))}{q!}\ \bra\ffi,\,y^q\ffi\ket(t)\
c_{N-n,p,l,k,\beta}(w,t)\,\right\|
\\[5pt]\nonumber
&\leq&\frac{c_6\,D_2^{n+3}\,(d_0 \|A\|)^{n+3}}{(1+(n+3))^{d+1}}\
\frac{\sqrt{(J+3+N+2(p-k-|l|))!}}{\sqrt{J!}}\
D_1\ D_2^{|l|+4(N-n)}\ l!\ \frac{|t|^p}{p!}\ \frac{k^k}{\delta^k}.
\eea
Then we use similar estimates to the above and the restrictions (\ref{restr})
to get
\bea
&&k^{2k}\,l!\,l!\,\frac{(J+3+N+2(p-k-|l|))!}{p!\,p!}\ \leq\
(2k)^{2k}\ \frac{(J+3+N+2(p-k))!}{p!\,p!}\nonumber\\[4pt]\nonumber
&\leq&(2k)^{2k}\ \frac{(J+3+3N-2n-2k)!}{(N-n)!\,(N-n)!}\ \leq\
\frac{(J+3+3N-2n)^{J+3+3N-2n}}{(N-n)!\ (N-n)!}.
\eea
Using this and $|l|\leq 3(N-n)/2$, we see that (\ref{1064}) is bounded above by
$$
\frac{c_6 D_1 (D_2 d_0 \|A\|)^{3}}{4^{d+1}\sqrt{J!}}\ D_2^{11 N/2}\ \,
\frac{(d_0 \|A\|)^{n}}{D_2^{9n/2}}\ \frac{|t|^p}{(N-n)!\delta^k}\
(J+3+3N-2n)^{(J+3+3N)/2}.
$$
Finally, with
$$
\sum_{p=0}^{N-n}\ \,\sum_{k=0}^{p+\frac{N-m}{2}}\ |t|^p\,\delta^{-k}\ \leq\
K_1\,K_2\,\delta^{-(N-m)/2}\
\left(\frac{t}{\delta}\right)^{N-n},
$$
(see (\ref{45}), (\ref{46})), the bounds $\sum_{|l|\leq p+(N-n)/2}\,1\,\leq\,
\sigma_0\,\E^{3\sigma(N-n)/2}$,\ \,
$\sum_{|q|\leq n+3}\,1\,\leq\,\sigma_0\,\E^{\sigma(n+3)}$,
and $|{\cal B}_{1, N-n}|\leq \E^{K_0(N-n)}$, we get
(with the conditions (\ref{restr}) on the summations)
\bea\label{yeah}
&&\sum_{|q|=2+n+1}\ \sum_{\beta,p,k,l,j}\,\eps^{N}\ \left\|\,
\frac{D^q\,E(\zeta_q(y,t,\eps))}{q!}\ \bra\ffi,\,y^q\ffi\ket(t)\,
c_{N-n,p,l,k,\beta}(w,t)\,\right\| \nonumber\\[5pt]
&\leq&\frac{\sigma_0^2 \E^{3\sigma} K_1K_2c_6 D_1 (D_2 d_0 \|A\|)^{3}}{4^{d+1}
\sqrt{J!}}\,\E^{5 \sigma N/2}\,D_2^{N}\,(d_0\|A\|)^{N}\,\eps^{N}\,
(J+3+3N)^{(J+3+N)/2}\nonumber\\[5pt]
&&\qquad\qquad\qquad\quad\times\quad
\left(\frac{D_2^{9/2}}{\delta^{1/2}\,d_0\,\|A\|}\right)^{N-n}\
\left(\frac{t}{\delta}\right)^{N-n}\ \frac{(J+3+3N)^{N-n}}{(N-n)!}.
\eea
Postponing the study of the $t$ and $J$ dependence of our
estimates, we use the bound\\
$(J+3+3N)^{(J+3+N)/2}\,\leq\,N^{N/2}\,(J+3+3N)^{(J+3)/2}\,(J+6)^{N/2}$
to establish the existence of constants
$L_0, L_1, L_2$, independent of $N$ and $n$, such that (\ref{yeah}) is bounded
above by
$$
L_0\,L_1^N\,\eps^N\,N^{N/2}\,\frac{(L_2\,N)^{N-n}}{(N-n)!}.
$$
It remains for us to sum over $n$ and use (\ref{norma}) to bound (\ref{need})
by
\bea\nonumber
&&
\eps^3\,\sqrt{\sigma_0}\,\E^{3\sigma N/2}\,L_0\,L_1^N\,\eps^N N^{N/2}\
\sum_{n=0}^N\ \frac{(L_2\,N)^{N-n}}{(N-n)!}\\[4pt]\nonumber 
&\leq&\eps^3\,\sqrt{\sigma_0}\,L_0\,(\E^{3\sigma/2}\,L_1)^N\,\eps^N\,N^{N/2}\
\sum_{s=0}^{\infty}\ \frac{(L_2\,N)^{s}}{s!}\\[5pt]\nonumber
&\leq&\eps^3\,\sqrt{\sigma_0}\,L_0\,(\E^{3\sigma/2}\,L_1\,\E^{L_2})^N\,\eps^N
\,N^{N/2}.
\eea
If we choose $g < 1/(L_1\,\E^{L_2+3\sigma/2})$,
we can apply the analysis (\ref{analysis})
to obtain an exponentially small
bound on (\ref{need}) by the optimal truncation
$N(\eps)\,=\,\grintl g^2/\eps^2\grintr$.

Since the estimates we have on the $d$'s are similar to those we have
on the $c$'s, with the replacement of $n$ by $n-1$, the same
exponential bound is valid for (\ref{sumerr2}), (see (\ref{normb}))
and the analysis of the the first collection of error terms
is completed.

\vskip .3cm
We now need to take into account the error terms
(\ref{cutoff1}) to (\ref{cutoff2}) arising from the derivatives of
the cut-off function $F$. Choose $F_0>0$ that satisfies
$$
\max\ \{\,|\Delta_w F(w)|,\ \|\nabla_w F(w)\|\,\}\ \leq\ F_0,
$$
uniformly in $w$, and recall that for any $i=1,\dots,d$,
\be
\mbox{supp}\,\partial_{w_i}F(w)\,\subseteq\,\{w\in\R^d\;:\; b_0<|w|<b_1\}
\ee
for some $0<b_0<b_1<\infty$.
Now consider (\ref{cutoff1}). We express $g_n$ in terms of the $c$'s to see that
the norm of $\ds\frac{2}{\eps^4}$ times (\ref{cutoff1})
(in $L^2(\R^d,{\cal H}_{\mbox{\scriptsize el}}) $) can be
bounded as follows:
\bea
&&\left\|\,\sum_{n=0}^N\ \Delta_w\,F(w)\,g_n(w,y,t)\,\eps^n\,\Phi(w,t)\,\right\|
\nonumber\\[4pt]
&\leq&\sum_{n=0}^N\ \eps^n\ \sqrt{\int_{\smallR^d}\,\left|\,\Delta_wF(w)
\sum_{|j|\leq J+3n}c_{n,j}(w,t)\,\eps^{-d/2}\,\ffi_j(w/\eps, t)\,\right|^2
\|\Phi(w, t)\|^2_{{\cal H}_{\mbox{\scriptsize el}}}\,dw}\nonumber\\
&\leq&F_0\,\sum_{n=0}^N\,\eps^n\,\sup_{w\in {\mbox{\scriptsize supp}F \subseteq \smallR^d}}\|c_n(w,t)\|\
\sqrt{\sum_{|j|\leq J+3n}\,\int_{|w|\geq b_0}\,
\left|\eps^{-d/2}\,\ffi_j(w/\eps,t)\right|^2\,dw}.
\eea
We know from Section 7 of \cite{hagjoy5} that there exists a constant
$0<\beta_d$ depending on the dimension $d$ only, such that
$$
\sqrt{2|j|+d}\ <\ b_0/(\| A\|\eps),\qquad\mbox{for all}\ |j|\leq J+3N
$$
and $|j|\leq J+3N$ imply
$$
\sqrt{\,\int_{|w|\geq b_0}\,\left|\eps^{-d/2}\,\ffi_j(w/\eps, t)\right|^2\,dw}
\ \leq\ \E^{\beta_d|j|}\ \E^{-\,(b_0^2)/(12\,\|A\|^{2}\,\eps^2)}.
$$
All the conditions here will be satisfied if
$N(\eps)\,=\,\grintl g^2/\eps^2\grintr$, provided
we choose $g$ and $\eps$ to satisfy
$$
\eps^2(d+2J)+6g^2 < b_0^2/\| A\|^2.
$$
For such a choice, using
$\sum_{|j|\leq J+3n}\,\E^{2\,\beta_d\,|j|}\,\leq\,
\sigma_0\,\E^{(\sigma+2\beta_d)(J+3n)}$,
we get
$$
\sqrt{\,\sum_{|j|\leq J+3n}\,\int_{|w|\geq b_0}
\left|\eps^{-d/2}\ffi_j(w/\eps, t)\right|^2\, dw}
\ \leq\ \sqrt{\sigma_0}\ \E^{(\sigma+2\beta_d)(J+3n)/2}
\ \E^{-\,(b_0^2)(12\|A\|^{2}\eps^2)}.
$$
Moreover, by means of manipulations that by now are familiar,
\bea\nonumber
&&\|c_n(w,t)\|\ \leq\ \sum_{\beta\in{\cal B}_{n,1}}\ \sum_{p\leq n}\
\sum_{k+|l|\leq p+\frac n2}\
\left\|c_{n,p,l,k,\beta}(w,t)\right\|\\[5pt]\nonumber
&\leq&\sum_{\beta\in {\cal B}_{n,1}}\ \sum_{p\leq n}\
\sum_{k+|l|\leq p+\frac n2}\ D_1\,
D_2^{|l|+4n}\,l!\,\frac{|t|^p}{p!}\ \frac{k^k}{\delta^k}\
\frac{(J+n+2(p-|l|-k))}{\sqrt{J!}}\\[5pt]\nonumber
&\leq&\sum_{\beta\in{\cal B}_{n,1}}\ \sum_{p\leq n}\
\sum_{k+|l|\leq p+\frac n2}\  \frac{D_1 D_2^{11n/2}}{\sqrt{J!}}\
\frac{|t|^p}{\delta^k}\ \frac{(J+3n)^{(J+3n)/2}}{n!}\\[5pt]\nonumber
&\leq&\E^{K_0n}\,\sigma_0\,\E^{3\sigma n/2}\,K_1 K_2\,\frac{1}{\delta^n}
\,\left(\frac{|t|}{\delta}\right)^n\,\frac{D_1 D_2^{11n/2}}{\sqrt{J!}}\,
(J+3n)^{J/2}\,\frac{(J+3N)^{3n/2}}{n!}.
\eea
Combining these estimates, we get the existence of positive constants
$M_0$ and $M_1$, such that for $N\,=\,\grintl g^2/\eps^2\grintr$,
\bea
&&\left\|\,\sum_{n=0}^N\,\Delta_w\,F(w)\,g_n(w,y,t)\,\eps^n\,\Phi(w,t)\,\right\|
\ \leq\ \E^{-\,(b_0^2)/(12\,\|A\|^{2}\,\eps^2)}\ \sum_{n=0}^N\,M_0\,
\frac{(\eps M_1 N^{3/2})^n}{n!}\nonumber\\[5pt]\nonumber
&\leq&\E^{-\,(b_0^2)/(12 \|A\|^{2}\eps^2)}\,M_0\,\E^{\eps M_1 N^{3/2}}
\ \leq\ \E^{-\,(b_0^2)/(12 \|A\|^{2}\eps^2)}\,M_0\,\E^{M_1 g^3/\eps^2}
\ \leq\ M_0\,\E^{-\,(b_0^2)/(24 \|A\|^{2}\eps^2)},
\eea
provided
$$
  M_1 g^3<b_0^2/(24 \|A\|^{2}).
$$
All other terms in the list (\ref{cutoff1}) to (\ref{cutoff2})
can be estimated in a similar fashion under a similar condition
on $g$.

This concludes the proof of our lemma.\ep

\vspace{.5cm}
\noindent{\bf Remark:}\quad
It is not difficult to check that if we keep $N$ fixed, then
our approximation (\ref{approx}) $\hat{\psi}(w,y,t)$ is accurate up to an error
of order $\eps^N$, as expected.

\vspace{.3cm}
A by--product of our estimates on the terms stemming from
the introduction of the cutoff is that our approximation is
exponentially localized in a ball centered at $a(t)$ of any radius
$b_0$, as stated in the second part of Theorem \ref{main}.

Hence, we have completed the proof of Theorem \ref{main}.\ep

\vskip .5cm
\section{Generalizations}\label{gene}
\setcounter{equation}{0}

As in \cite{hagjoy5}, under some mild supplementary assumptions,
we can extend our results to allow $0\leq t\leq T(\eps)$
with $T(\eps)\simeq \ln(1/\eps^2)$. This proves the validity of our
construction up to the Ehrenfest time scale.

\begin{thm}\label{ehr}
In addition to the assumptions of Theorem \ref{main},
assume that a classical solution to the equation
(\ref{newton}) exists for all $t\in\R$. Moreover, assume that
for all $z$ in a complex neighborhood of $\Xi$, the following bound is satisfied
$$
|E(z)|\ \leq\ N\,\E^{M|z|},
$$
and that $E(x)$ is bounded below.
Suppose also that there exist $L$ and $\lambda>0$, such that for all $t\in\R$
$$
\|A(t)\|\,+\,\|B(t)\|\ \leq\ L\,\E^{\lambda t}.
$$
Then, there exist $\tau'$, $C'$, $T'>0$, and $0<\sigma,\,\sigma'<2$ such that
the approximation defined by choosing $N(\eps)\simeq 1/\eps^{\sigma}$
is accurate up to an error whose norm is bounded by
$C'\E^{-\tau'/\eps^{\sigma'}}$,
uniformly for all times $0\leq t\leq T'\,\ln(1/\eps^2)$.
\end{thm}

{\bf Proof:}\quad It is enough to mimick the proof of the corresponding result
for the semiclassical propagation of the Schr\"odinger equation in
\cite{hagjoy5}, since
our hypotheses imply that nothing can happen on the adiabatic side of the
problem. By the conservation of energy, the exponential bound on $E(z)$ and
the assumed existence of a Liapunov exponent, we easily see from the proof of
Lemmas \ref{nbehave} and \ref{optrunc}, that the behavior in $t$ of all
constants (independent of $N$) is at worst exponential in $t$.
From the conditions $D_2\geq \E^{K T}$, with $K$ some constant, we
need to take $g(T)\leq g_0\E^{-g_1 t}$ so that the optimal truncation
procedure yields an error of the order
$\ds\E^{K_0 T}\E^{-g_0^2\e^{-2g_1}/\eps^2}$. The choice $T(\eps)\leq
T'\ln(1/\eps^2)$, with $T'>0$ sufficiently small, gives the desired result.\ep

\vspace{.3cm}
Similarly, we can extend our results to allow initial conditions
in a wider class of vectors. Indeed, we have been careful
to make explicit the $J$ dependence in all estimates so that we
can control the error term as a function of $J$. Recall that
$J$ is fixed arbitrarily in (\ref{2aorder}) which gives the
expansion in the basis $\ffi_j(A(0),B(0),\eps^2,a(0),\eta(0),x)$
of the nuclear part of the wave function that we take as an initial condition.

\vspace{.2cm}
As in \cite{hagjoy5}, for $(a,\eta)\in\R^{2d}$, we introduce
the operator $\Lambda_{\eps}(a,\eta)$ such that
$$
(\Lambda_{\eps}(a,\eta)f)(x)\,=\,
{\eps}^{-d}\,\E^{i\eta \cdot\,(x-a)/{\eps^2}}\,f((x-a)/{\eps}).
$$
We define a dense set ${\cal C}$ in $L^2(\R^d)$, that is contained in the set
${\cal S}$ of Schwartz functions, by
\bea
{\cal C}&=&\left\{\,f(x)\,=\,\sum_{j}\,c_j\,\ffi_j(\un,\,\un,\,1,\,0,\,0,\,x)\,
\in\,{\cal S},\right.\mbox{ such that }\nonumber\\
&&\qquad\qquad\quad\left.\,\mbox{there exists}\ K>0 \mbox{ with }
\sum_{|j|>J}|c_j|^2\leq
\E^{-K J}, \mbox{ for large } J\,\right\}.\label{woofer}
\eea

\vskip .2cm \noindent
{\bf Remark}\quad It is easy to check that the inequality in (\ref{woofer}) is
equivalent to the requirement that the coefficients of $f$ satisfy
$$
|c_j|\,\leq\,\E^{-K|j|},
$$
for large $|j|$.
Another equivalent definition of ${\cal C}$ is
$$
{\cal C}\,=\,\cup_{t>0}\ \E^{-tH_{ho}}\,{\cal S},
$$
where $H_{ho}\,=\,-\,\Delta/2\,+\,x^2/2$ is the harmonic oscillator
Hamiltonian. The set ${\cal C}$ is also called the set of analytic vectors
\cite{RS} for the harmonic oscillator Hamiltonian.

\vskip .4cm
Let $f\in{\cal C}$. We set
\bea
f_J(y, t) &=& \sum_{|j|\leq J}\,
c_j\,\ffi_j(A(t),B(t),\eps^2,0,0,y),
\qquad\mbox{and}\nonumber\\[5pt]
f(y, t) &=& \sum_{j}\,c_j\,\ffi_j(A(t),B(t),\eps^2,0,0,y)\nonumber
\eea
where the classical quantities $a(t)$, $\eta(t)$, $A(t)$, $B(t)$, and $S(t)$
correspond to the initial conditions
$a(0)$, $\eta(0)$, $A(0)=B(0)=\un$, and $S(0)$.
We consider the construction described in Section \ref{mr} corresponding
to the initial condition $g_0(0,y,t)=f_J(y,t)$, making explicit the dependence
on $J$ in the notation:
\bea\nonumber
&&\hat{\Psi}_{J, N}(w, y, t)\\
&=&F(w)\,\E^{iS(t)/\eps^2}\,\E^{i\eta(t)\cdot y/\eps}\,
\left(\sum_{n=0}^N\,\eps^n\,g_{n,J}(w,y,t)\,\Phi(w,t)\,+
\sum_{n=2}^{N+2}\,\eps^n\,
\phi_{n,J}^\perp(w,y,t)\right).\nonumber
\eea
Recall that
\bea\nonumber
&&\hat{\Psi}_{J,N}(w,y,0)\\
&=&F(w)\,\E^{iS(0)/\eps^2}\,\E^{i\eta(0)\cdot y/\eps}\,
\left(\,f_J(y, 0)\,\Phi(w, 0)\,+\,
\sum_{n=2}^{N+2}\,\eps^n\,
\phi_{n, J}^\perp(w,y,0)\,\right).\nonumber
\eea

Let $\nu>0$, and consider $N(\eps)\,=\,\grintl g^2/\eps^2\grintr$
and $J(\eps)=\nu N(\eps)$.
We define our more general initial conditions as
\bea\nonumber
&&\hat{\Psi}_{f}(w,y,0)\\
&=&F(w)\,\E^{iS(0)/\eps^2}\,\E^{i\eta(0)\cdot y/\eps}\,
\left(f(y,0)\,\Phi(w, 0)\,+\,
\sum_{n=2}^{N(\eps)+2}\,\eps^n\,
\phi_{n,J(\eps)}^\perp(w,y,0)\right),\nonumber
\eea
which corresponds, when we get back to the variables $(X,t)$, to an initial
state $\hat{\Psi}_{f}(X-a(0),(X-a(0))/\eps,0)$ whose projection along
the electronic eigenvector $\tilde{\Phi}(X,0)$ yields a nuclear wave packet of
the form $(\Lambda_{\eps}(a(0),\eta(0))f)(X)$.
Note that the component of the initial state
perpendicular to $\tilde{\Phi}(X,0)$ necessary to achieve exponential
accuracy depends on $\eps$.
This component is determined by the coefficients of the function $f$.

We can now state our result for such general initial conditions
\begin{thm}
Assume the hypotheses of Theorem \ref{main} and consider the above
constructions. There exist sufficiently small $g>0$ and positive constants
$C(g)$, $\Gamma(g)$, such that with the definition
$$
\Psi_*(X,t,\eps)=\hat{\Psi}_{J(\eps),N(\eps)}(X-a(t),(X-a(t))/\eps,t),
$$
we have
$$
\left\|\,\E^{-itH(\eps)/\eps^2}\,\Psi_f(X, 0, \eps)\,-\,\Psi_*(X,t,\eps)\,
\right\|_{L^2({\smallR}^d,{\cal H}_{\mbox{\scriptsize el}})}
\ \leq\ C(g)\,\E^{-\Gamma(g)/\eps^2},
$$
for all $t\in [0, T]$, as $\eps\ra 0$.\\
Moreover, the result for times $T\simeq \ln(1/\eps^2)$
corresponding to Theorem \ref{ehr} is also true for these initial
conditions.
\end{thm}

\noindent{\bf Proof:}\quad We have
\bea\nonumber
&&\E^{-itH(\eps)/\eps^2}\,\Psi_f(X,0,\eps)\\[7pt]\nonumber 
&=&\E^{-itH(\eps)/\eps^2}\,
(\Psi_f(X,0,\eps)-\Psi_*(X,0,\eps))\,+\,\E^{-itH(\eps)/\eps^2}\,
\Psi_*(X, 0,\eps)
\\[7pt]\nonumber
&=&\Psi_*(X,t,\eps)\ +\
O(\|\E^{-itH(\eps)/\eps^2}\,\Psi_*(X,0,\eps)-\Psi_*(X,t,\eps)
\|_{L^2({\smallR}^d,{\cal H}_{\mbox{\scriptsize el}})})\\[7pt]
\nonumber
&&+\quad O\left(\,\|\Psi_f(X,0,\eps)-\Psi_*(X, 0, \eps)
\|_{L^2({\smallR}^d,{\cal H}_{\mbox{\scriptsize el}})}\,\right).
\eea
By our choice of function $f$, the last term is exponentially small
in $1/\eps^2$. The remaining norm to estimate corresponds to the situation
of Theorem \ref{main} in which we let the parameter $J$ grow as
$1/\eps^2$, according to our choice of $J(\eps)$. But, as
in the proof of Theorem 3.6 in \cite{hagjoy5} for the corresponding
result in semiclassical dynamics, we have made the dependence in $J$
of all the key estimates explicit. It is enough to go through the proof of
theorem \ref{main} to check that with
$J=\nu N$, all arguments can be repeated to get the same $N$ and $\eps$
behavior for the estimates on the error terms, (see \cite{hagjoy5} for details).
Hence, we see that for sufficiently small $g$, we can approximate the solution
corresponding to these generalized initial conditions up to an error of
order $\E^{-\Gamma(g)/\eps^2} $. The Ehrenfest time regime is dealt
with similarly. \ep

\vskip .5cm
\section{Technicalities}\label{techie}
In this section we give the proofs of the auxiliary lemmas we used in the
course of the main argument.

\vskip .5cm
\noindent
{\bf Proof of Lemma \ref{cauchylem}:}\quad We first consider the case $k\ge 1$.
By Cauchy's formula, we can write
\be\label{cauchyformula}
g'(t)\ =\ \frac 1{2\pi i}\ \int_\Gamma\,\frac{g(s)}{(t-s)^2}\,ds,
\ee
where $\Gamma$ is the circular contour with center $t$ and radius
$\ds\frac 1{k+1}\ (\delta\,-\,|\mbox{\rm Im}\,t|)$.

For $s$ on $\Gamma$, we have
$\ds (\delta-|\mbox{\rm Im}\,s|)\,\ge\,
\frac k{k+1}(\delta\,-\,|\mbox{\rm Im}\,t|)$. Thus,
$$
\|g(s)\|\quad\le\quad C\ k^k\ (\delta\,-\,|\mbox{\rm Im}\,s|)^{-k}
\quad\le\quad C\ k^k\ \left[\,\frac{k}{k+1}\,
(\delta\,-\,|\mbox{\rm Im}\,t|)\,\right]^{-k}
$$
So, by putting the norm inside the integral in (\ref{cauchyformula}), we have
\bea\nonumber
\|g'(t)\|&\le&\frac 1{2\pi}\
\frac{2\pi}{k+1}(\delta\,-\,|\mbox{\rm Im}\,t|)\
C\,k^k\,\left[\frac{k}{k+1}(\delta\,-\,|\mbox{\rm Im}\,t|)\right]^{-k}\
\left[\frac{1}{k+1}(\delta\,-\,|\mbox{\rm Im}\,t|)\right]^{-2}\\[5pt]\nonumber
&=&C\ (k+1)^{k+1}\ (\delta\,-\,|\mbox{\rm Im}\,t|)^{-k-1}.\eea

For $k=0$ we use the same argument with the radius of $\Gamma$ replaced by
$\ds\alpha\,\,(\delta\,-\,|\mbox{\rm Im}\,t|)$ for any $\alpha<1$. This yields the bound
$$\|g'(t)\|\ \le\ C\,\alpha^{-1}\,(\delta\,-\,|\mbox{\rm Im}\,t|)^{-1}.$$
The lemma follows because $\alpha<1$ is arbitrary.\qquad\qquad\ep

\vskip .5cm \noindent
{\bf Proof of Lemma \ref{nenciuprop2}}:\quad
To prove the quantity $\nu$ is finite, we estimate
\bea\nonumber
&&\sum_{\{\,l\,:\,0\le l_i\le\alpha_i\,\}}\
\frac{1}{(1+|l|)^{d+1}}\ \frac{1}{(1+|\alpha-l|)^{d+1}}\\[8pt] \nonumber
&=&\sum_{\begin{array}{c}
{\{\,l\,:\,0\le l_i\le\alpha_i\,\}}\\[2pt]
|l|\le\grintl\frac{|\alpha|}2\grintr\end{array}}\
\frac{1}{(1+|l|)^{d+1}}\ \frac{1}{(1+|\alpha-l|)^{d+1}}\\[-3pt] \nonumber
&&\quad\quad\qquad\qquad\qquad\qquad\ +\qquad
\sum_{\begin{array}{c}
{\{\,l\,:\,0\le l_i\le\alpha_i\,\}}\\[2pt]
|l|>\grintl\frac{|\alpha|}2\grintr\end{array}}\
\frac{1}{(1+|l|)^{d+1}}\ \frac{1}{(1+|\alpha-l|)^{d+1}}\\[8pt] \nonumber
&\le&\frac{2}{\left(\,1+\grintl\frac{|\alpha|}2\grintr\,\right)^{d+1}}\
\sum_{\begin{array}{c}
{\{\,l\,:\,0\le l_i\le\alpha_i\,\}}\\[2pt]
|l|\le\grintl\frac{|\alpha|}2\grintr\end{array}}\
\frac{1}{(1+|l|)^{d+1}}\\[8pt] \nonumber
&\le&\frac{2^{d+2}}{\left(\,1+|\alpha|\,\right)^{d+1}}\
\sum_{\begin{array}{c}
{\{\,l\,:\,0\le l_i\le\alpha_i\,\}}\\[2pt]
|l|\le\grintl\frac{|\alpha|}2\grintr\end{array}}\
\frac{1}{(1+|l|)^{d+1}}\\[8pt] \nonumber
&\le&\frac{2^{d+2}}{\left(\,1+|\alpha|\,\right)^{d+1}}\
\sum_{l}\ \frac{1}{(1+|l|)^{d+1}}.
\eea

\vskip .2cm
\noindent
Thus,\qquad $\ds\nu\ \le\ 2^{d+2}\ \sum_l\ (1+|l|)^{-d-1}$.

\vskip .1cm
\noindent
To see that the right
hand side of this inequality is finite, we note that the number of
multi-indices $l$ with $|l|=L$ is the binomial coefficient
$\ds\left(\begin{array}{c}L+d-1\\d-1\end{array}\right)^{\phantom{|}}$,
with the convention that
$\ds\left(\begin{array}{c}0\\0\end{array}\right)\ =\ 1$.
Thus,
\bea\nonumber
\nu&\le&2^{d+2}\ \sum_{L=0}^\infty\
\left(\begin{array}{c}L+d-1\\d-1\end{array}\right)\ \frac{1}{(1+L)^{d+1}}\\[8pt]
&=&\frac{2^{d+2}}{(d-1)!}\ \sum_{L=0}^\infty\
\frac{(L+d-1)(L+d-2)\cdots(L+1)}{(L+1)^{d+1}}.\nonumber
\eea

\vskip .1cm \noindent
For large $L$,\quad $\ds\frac{(L+d-1)(L+d-2)\cdots(L+1)}{(L+1)^{d+1}}\ $
is asymptotic to $L^{-2}$, so $\nu$ is finite.

\vskip .3cm
\noindent
Since
$\ds D^\alpha\,(M\,N)\ =\
\sum_{\{\,l\,:\,0\le l_i\le\alpha_i\,\}}\
\left[\,\prod_{j=1}^d\,
\left(\begin{array}{c}\alpha_j\\l_j\end{array}\right)\,\right]\
\left(\,D^l\,M\,\right)\ \left(\,D^{(\alpha-l)}\,N\,\right)$, we have

\vskip .2cm
\bea\nonumber
&&\left\|\,\left(\,D^\alpha\,(M\,N)\,\right)(x)\,\right\|\\[6pt] \nonumber
&\le&
\sum_{\{\,l\,:\,0\le l_i\le\alpha_i\,\}}\,
\left[\,\prod_{j=1}^d\,
\left(\begin{array}{c}\alpha_j\\l_j\end{array}\right)\,\right]\
m(x)\ n(x)\ a(x)^{|\alpha+p+q|}\quad\frac{(l+p)!}{(1+|l|)^{d+1}}\
\frac{(\alpha-l+q)!}{(1+|\alpha-l|)^{d+1}}\\[7pt]
\nonumber&=&
m(x)\ n(x)\ a(x)^{|\alpha+p+q|}\quad(\alpha+p+q)!\\[6pt] \nonumber
&&\ \times\quad
\sum_{\{\,l\,:\,0\le l_i\le\alpha_i\,\}}
\left[\,\prod_{j=1}^d\,
\left(\begin{array}{c}\alpha_j\\l_j\end{array}\right)\,
\left(\begin{array}{c}\alpha_j+p_j+q_j\\l_j+p_j\end{array}\right)^{-1}\,\right]\
\frac 1{(1+|l|)^{d+1}\ (1+|\alpha-l|)^{d+1}}.
\eea
Since
$\ds\left(\begin{array}{c}\alpha_j+p_j+q_j\\l_j+p_j\end{array}\right)\ \ge\
\left(\begin{array}{c}\alpha_j+q_j\\l_j\end{array}\right)\ \ge\
\left(\begin{array}{c}\alpha_j\\l_j\end{array}\right)$,\quad we therefore have
\bea\nonumber
\left\|\,\left(\,D^\alpha\,(M\,N)\,\right)(x)\,\right\|
&\le&m(x)\ n(x)\ a(x)^{|\alpha+p+q|}\quad (\alpha+p+q)!\\[6pt] \nonumber
&&\qquad\qquad\qquad\times\qquad\sum_{\{\,l\,:\,0\le l_i\le\alpha_i\,\}}\
\frac 1{(1+|l|)^{d+1}\ (1+|\alpha-l|)^{d+1}}.\\[6pt] \nonumber
&\le&m(x)\ n(x)\ \nu\ a(x)^{|\alpha+p+q|}\quad
\frac{(\alpha+p+q)!}{(1+|\alpha|)^{d+1}}.\qquad\qquad\ep
\eea

\vskip .5cm \noindent
{\bf Proof of Lemma \ref{integrate}:}\quad
If $f(t)$ satisfies $\|f(t)\|\leq C\,|t|^p\,\dist(t)^{-k}$, for all
$t\in\Omega$,
there exists $g(t)$ analytic in $\Omega$, such that $f(t)=t^p\,g(t)$ and
$\|g(t)\|\leq C\,\dist(t)^{-k}$.
We use the integration path from $0$ to $t\in\Omega$ parametrized
by $\gamma(u)=t u$, with $u\in [0,1]$, to compute
\bea
&&\left\|\,\int_0^t\,f(s)\,ds\,\right\|\quad =\quad
\left\|\,\int_0^1\,f(tu)\,du\,\right\|
\quad =\quad \left\|\,\int_0^1\,t\,(tu)^p\,g(tu)\,du\,\right\|\nonumber\\[4pt]
&\leq&C\,|t|^{p+1}\,\int_0^1\,\frac{u^p}{\dist(tu)^{k}}\,du\
\quad\leq\quad C\ \frac{|t|^{p+1}}{p+1}\ \dist(t)^{-k},
\eea
since, by assumption, $\dist(ut)$ is a decreasing function of $u$.\ep


\begin{thebibliography}{xxxxxxx}
%
\bibitem{benchaou} Benchaou, M.:\quad Estimations de Diffusion
pour un Op\'erateur de Klein-Gordon Matriciel D\'ependant du Temps.
{\it Bull. Soc. math. France} {\bf 126},
273--294 (1998).
%
\bibitem{benmar} Benchaou, M., and Martinez A.:\quad Estimations
Exponentielles en Th\'eorie de la Diffusion des Op\'erateurs de
Schr\"odinger Matriciels. {\it Ann. Inst. H. Poincar\'e Sect. A} {\bf 71},
561--594 (1999).
%
\bibitem{lots!a} Berry, M.V.:\quad Quantum Phase Corrections from
Adiabatic Iteration. {\it Proc. R. Soc. Lond. A} {\bf 414}, 31--46
(1987).
%
\bibitem{lots!b} Berry, M.V.:\quad
Histories of Adiabatic Quantum Transitions.
{\it Proc. R. Soc. Lond. A} {\bf 429}, 61--72 (1990).
%
\bibitem{coker} Coker, D. F., and Xiao, L:\quad Methods for
Molecular--Dynamics with Nonadiabtic Transitions.
{\it J. Chem. Phys.} {\bf 102}, 496--510 (1995).
%
\bibitem{dieudonne} Dieudonn\'e J.:\quad {\it Calcul Infinit\'esimal}.
Paris: Hermann 1968.
%
%
\bibitem{oldstuff1} Hagedorn, G. A.:\quad A Time--Dependent
Born--Oppenheimer Approximation. {\it Commun. Math. Phys.} {\bf 77}, 1--19
(1980).
%
\bibitem{oldstuff2} Hagedorn, G. A.:\quad
High Order Corrections
to the Time--Dependent Born--Oppenheimer Approximation I: Smooth
Potentials.
{\it Ann. Math.} {\bf 124}, 571--590 (1986).
Erratum {\bf 126}, 219 (1987).
%
\bibitem{oldstuff3} Hagedorn, G. A.:\quad
High Order Corrections to
the Time--Dependent Born--Oppenheimer Approximation II:
Coulomb Systems.
{\it Commun. Math. Phys.} {\bf 117}, 387--403 (1988).
%
\bibitem{masterpiece} Hagedorn, G. A.:\quad Molecular Propagation Through
Electronic Eigenvalue Crossings,
{\it Memoirs Amer. Math. Soc.} {\bf 536}, (1994).
%
\bibitem{semi3} Hagedorn, G.\,A.:\quad {Semiclassical Quantum Mechanics
III: The Large Order Asymptotics and More General States.}
\textit{Ann. Phys.} \textbf{135}, 58--70 (1981).
%
\bibitem{semi4} Hagedorn, G.\,A.:\quad {Semiclassical Quantum Mechanics IV:
Large Order Asymptotics and More General States in More than One Dimension.}
\textit{Ann. Inst. H. Poincar\'e Sect. A.} \textbf{ 42}, 363--374 (1985).
%
\bibitem{raise} Hagedorn, G.\,A.:\quad Raising and lowering operators for
semiclassical wave packets. {\it Ann. Phys.} {\bf 269}, 77--104 (1998).
%
\bibitem{hagjoy3} Hagedorn, G.\,A. and Joye, A.:\quad Semiclassical Dynamics
with Exponentially Small Error Estimates. {\it Commun. Math. Phys.}
\textbf{207}, 439--465 (1999).
%
\bibitem{hagjoy5} Hagedorn, G. A., and Joye, A.:\quad Exponentially
Accurate Semiclassical Dynamics: Propagation, Localization, Ehrenfest
Times, Scattering and More General States.
{\it Annales Henri Poincar\'e}\ \,(to appear).
%
\bibitem{j} Joye, A.:\quad Proof of the Landau--Zener
Formula. {\it Asymptotic Analysis} {\bf 9}, 209--258 (1994).
%
\bibitem{jp} Joye, A. and  Pfister, C.-E.:\quad Exponentially Small
Adiabatic Invariant for the Schr\"odinger Equation. 
{\it Commun. Math. Phys.} {\bf 140}, 15--41 (1991).
%
\bibitem{alainetal1} Joye, A. and  Pfister, C.-E.:\quad Superadiabatic
Evolution and Adiabatic
Transition Probability between Two Non--Degenerate Levels Isolated
in the Spectrum. {\it J. Math. Phys.} {\bf 34}, 454--479 (1993).
%
\bibitem{alainetal2} Joye, A., Pfister, C.-E. :\quad
Semi-Classical Asymptotics beyond All Orders
for Simple Scattering Systems, {\em SIAM J. Math. Anal.} {\bf 26},
944--977 (1995).
%
\bibitem{h30} Klein, M.:\quad On the Mathematical Theory of
Predissociation. {\it Ann. Phys.} {\bf 178}, 48--73 (1987).
%
\bibitem{lots!2} Lim R., and Berry, M.V.:\quad
Superadiabatic Tracking of Quantum Evolution.
{\it J. Phys. A: Math. Gen.} {\bf 24}, 3255--3264 (1991).
%
\bibitem{h34}
Martinez, A.:\quad D\'eveloppements Asymptotiques et Effet
Tunnel dans l'Ap\-proximation de Born--Oppenheimer. {\it Ann. Inst. H.
Poincar\'e Sect. A} {\bf 50}, 239--257 (1989).
%
\bibitem{h36} Martinez, A.:\quad Resonances dans
l'Approximation de Born--Oppenheimer I. {\it J. Diff. Eq.}
{\bf 91}, 204--234 (1991).
%
\bibitem{h37}  Martinez, A.:\quad Resonances dans
l'Approximation de Born--Oppenheimer II. Largeur de R\'esonances.
{\it Commun. Math. Phys.} {\bf 135}, 517--530 (1991).
%
\bibitem{nenciu} Nenciu, G.:\quad Linear Adiabatic Theory and
Applications: Exponential Estimates. {\it Commun. Math. Phys.}
{\bf 152}, 121--135 (1993).
%
\bibitem{pechukas} Pechukas, P.:\quad Time--Dependent Semiclassical
Scattering Theory.\quad II. Atomic Collisions. {\it Phys. Rev.}
{\bf 181}, 174--184 (1969).
%
\bibitem{RS} Reed, M. and Simon, B.:\quad {\it Methods of Modern
Mathematical Physics
I:\ Functional Analysis}. New York, London: Academic Press 1972.
%
\bibitem{tully} Tully, J. C.:\quad Molecular Dynamics with Electronic
Transitions. {\it J. Chem. Phys.} {\bf 93}, 1061--1071 (1990).
%
\bibitem{webster} Webster, F., Rossky, P. J., and Friesner, R. A.:\quad
Nonadiabatic Processes in Condensed Matter: Semi--Classical Theory and
Implementation. {\it Comp. Phys. Commun.} {\bf 63}, 494--522 (1991).

%
\end{thebibliography}
\end{document}